\documentclass[12pt,a4paper]{article}
\usepackage{amsmath,amssymb,mathrsfs,framed,esint,slashed,upgreek}
\usepackage[colorlinks]{hyperref}
\usepackage{color}
\usepackage[all]{xy}

\usepackage{abstract}

\newtheorem{theorem}{Theorem}[section]

\newtheorem{remark}[theorem]{Remark}

\newcommand{\comment}[1]{\vspace{5mm}\par
\framebox{\begin{minipage}[c]{.95 \textwidth} \tt\bfi #1
\end{minipage}}\vspace{5 mm}\par}

\newcommand{\rem}[1]{}

\newcommand{\de}{{\rm d}}

\newcommand{\bq}{{\boldsymbol{q}}}
\newcommand{\bv}{{\boldsymbol{v}}}
\newcommand{\bp}{{\boldsymbol{p}}}

\newcommand{\bm}{{\mathbf{m}}}

\newcommand{\bM}{{\mathbf{M}}}
\newcommand{\bX}{{\mathbf{X}}}

\newcommand{\br}{{\boldsymbol{r}}}
\newcommand{\bx}{{\boldsymbol{x}}}

\newcommand{\bsigma}{{\boldsymbol{\sigma}}}

\newcommand{\bz}{{\mathbf{z}}}
\newcommand{\bw}{{\mathbf{w}}}

\newcommand{\bE}{{\mathbf{E}}}
\newcommand{\bB}{{\mathbf{B}}}

\newcommand{\bu}{{\boldsymbol{u}}}

\newcommand{\bGamma}{{\boldsymbol{\Gamma}}}

\newcommand{\bfi}{\bfseries\itshape}

\newcommand{\beq}{\begin{equation}}
\newcommand{\eeq}{\end{equation}}
\newcommand{\ben}{\begin{eqnarray}}
\newcommand{\een}{\end{eqnarray}}

\renewcommand{\contentsname}{}

\begin{document}

\title{\vspace{-.9cm}Complex fluid models\\of mixed quantum-classical dynamics}
\author{Fran\c{c}ois Gay-Balmaz$^1$, Cesare Tronci$^{2}$ 
\\ 
\footnotesize
\it $^1$Division of Mathematical Sciences, Nanyang Technological University, Singapore
\\
\footnotesize
\it $^2$Department of Mathematics, University of Surrey, Guildford, United Kingdom
}
\date{\vspace{-1.35cm}}

\maketitle

\begin{abstract}
Several methods in nonadiabatic molecular dynamics are based on Madelung's hydrodynamic description of nuclear motion, while the electronic component is treated as a finite-dimensional quantum system. In this context, the quantum potential leads to severe computational challenges and one often seeks to neglect its contribution, thereby approximating nuclear motion as classical. The resulting model couples classical  hydrodynamics for the nuclei to the quantum  motion of the electronic component, leading to the structure of a complex fluid system. This type of mixed quantum-classical fluid models have also appeared in solvation dynamics to describe the  coupling between liquid solvents and the quantum solute molecule. While these approaches  represent a promising direction, their mathematical structure  requires a certain care. In some cases, challenging higher-order gradients make these equations hardly tractable. In  other cases, these models are based on phase-space formulations that suffer from well-known consistency issues. Here, we present a new  complex fluid system that resolves these difficulties.
Unlike common approaches, the current system is obtained by applying the fluid closure at the level of the action principle of the original phase-space model. As a result, the system inherits a Hamiltonian structure and retains energy/momentum balance. After discussing some of its structural properties and dynamical invariants, we illustrate the model in the case of pure-dephasing dynamics. We conclude by presenting some invariant planar models.
\vspace{-.65cm}
\end{abstract}

{\footnotesize
\contentsname
\tableofcontents
}

\section{Introduction}

\subsection{Madelung hydrodynamics in  molecular motion\label{sec:QHD}}

The search for cost-effective computational approaches to many-body quantum systems has stimulated a great deal of studies, often exploiting Madelung's hydrodynamic formulation of quantum mechanics \cite{Madelung}. For example, the hydrodynamic approach has  led to the development of several trajectory-based algorithms in quantum molecular dynamics \cite{Chattaraj,Garashchuk10,HuPaBu09,Kendrick,Wyatt}, beyond the well-established Born-Oppenheimer theory. In this setting, one deals with the coupled motion of both nuclei and electrons in a molecular system for which the full Schr\"odinger equation represents a formidable challenge. The latter is normally tackled by convenient approaches that alleviate the computational complexity. 
Among these approaches, Madelung hydrodynamics has the great advantage of restoring the concept of trajectory, which can then be exploited to model the nuclear dynamics while the electronic motion is usually approximated as a finite-dimensional quantum system \cite{agostini2016quantum,Curchod1,HoRaTr21}.

In this hydrodynamic picture, the Lagrangian  trajectories associated to the nuclear motion sweep the electronic unitary evolution. In many important cases \cite{agostini2016quantum,HoRaTr21}, this description is made possible by the observation that any  two-particle wavefunction $\Psi(\br,\bx,t)$ may be expressed in terms of the \emph{exact factorization} \cite{abedi2012correlated}, i.e. $\Psi(\br,\bx,t)=\chi(\br,t)\psi(\bx,t;\br)$, with $|\chi|^2\neq 0$ and $\int|\psi|^2\,\de^3x=1$. Here, the semicolon indicates that $\psi$ is a  conditional wavefunction so that the coordinate $\br$ appears merely as a parameter. Notice that this factorization is directly generalizable to the case of two families of several particles, i.e., nuclei with coordinates $\{\br_j\}$ and electrons with coordinates $\{\bx_k\}$, although here we will deal the simplest possible case of only one nucleus. In addition, to avoid possible complications from functional analysis, we will consider finite-dimensional truncations of  the conditional wavefunction. This means that $\psi(\bx,t;\br)$ is replaced by a field $\psi(t;\br)$ taking values in the electronic Hilbert space  $\mathsf{H}=\Bbb{C}^n$.
 Replacing the exact factorization ansatz in the two-particle Schr\"odinger equation $i\hbar\partial_t\Psi=-\hbar^2\Delta_\br\Psi/M+\widehat{H}(\br)\Psi$ and writing $\chi=\sqrt{D}e^{-iS/\hbar}$, as well as $\hat{\varrho}=\psi\psi^\dagger$, yields the  system \cite{FoHoTr19,HoRaTr21}
\begin{align}
\begin{split}
&  M(\partial_t + \bu\cdot\nabla)\bu   = 
  - \nabla V_Q-\langle\hat{\varrho}|\nabla\widehat{H}\rangle -\frac{\hbar^2}{2MD}\partial_j\langle D \nabla\hat{\varrho}|  \partial_j\hat{\varrho}\rangle
  ,
  \\
  &
     i\hbar (\partial_t +\bu\cdot\nabla)\hat{\varrho} = \left[\widehat{H}-\frac{\hbar^2}{2MD}{\rm div}(D\nabla\hat{\varrho}),\hat{\varrho}\right]
,\qquad\quad
\partial_t D+{\rm div}(D\bu)=0\,.
\end{split}    \label{final-D-eqn}
\end{align}
Here, and throughout this paper, $\langle A |B\rangle=\operatorname{Tr} (A^\dagger B)$ is the Frobenius inner product on $\Bbb{C}^{n\times n}$ and the Hamiltonian operator $\widehat{H}(\br)$ is an $\br$-dependent Hermitian matrix. The latter is usually expressed as $\widehat{H}(\br)=V(\br)\boldsymbol{1}+\widehat{V}_I(\br)+\widehat{H}_Q$, where $V(\br)$ is  the scalar potential energy, $\widehat{H}_Q$ is an $\br$-independent matrix, and $\widehat{V}_I(\br)$ is a general interaction term. Also, we have denoted
\[
\bu:=\frac1MD\big(\nabla S-\langle\psi|i\hbar\nabla\psi\rangle\big)
,\qquad\qquad 
V_Q:= -\,\frac{\hbar^2}{2M}\frac{\Delta \sqrt{D}}{\sqrt{D}}.
\]
The density matrix of the electronic system  is expressed as $\hat\uprho=\int \Psi(\br)\Psi^\dagger(\br)\,\de^3r=\int D\hat\varrho\,\de^3r$. 
Equations \eqref{final-D-eqn} describe a system in which the quantum hydrodynamic flow sweeps a finite-dimensional quantum system, which  also undergoes its own unitary evolution. The latter comprises both a standard  term $[\widehat{H},\hat{\varrho}]$ of von Neumann type and an $\hbar^2$-term carrying second-order gradients. In turn, these terms also feed back in the hydrodynamic momentum equation. Together with the \emph{quantum potential} $V_Q$, these terms make the level of complexity of this system rather challenging. In particular, the quantum potential in Madelung hydrodynamics has motivated  relevant works in PDE analysis \cite{Gamba02,Gamba09,Degond1,Degond2}. 
We observe that \eqref{final-D-eqn} has the structure of a complex fluid model, in which a  macroscopic fluid equation is coupled to the evolution of a microscopic component. While in this case the latter is given by the density matrix of the quantum state, in liquid crystal theory this is given by the $Q$-tensor of the molecular rotational state \cite{deGennes1969}.
 In physical terms,  the quantum potential $V_Q$ is known to be responsible for the emergence of high-frequency and short-wavelength patterns that may  lead to the formation of regions of space where the density $D$ vanishes (quantum vortices) \cite{BialynickiBirulaSliwa2000,Dirac1931,FoTr23}. Even if one attempted to discard $V_Q$, the other nonlinear terms involving second-order gradients would still lead to analogous complications. Overall,  the full hydrodynamic system \eqref{final-D-eqn} is accompanied by computational difficulties that still challenge the  chemistry community
\cite{Garashchuk,GuFranco,Suzuki16,ZhMa03}. 

\subsection{Mixed quantum-classical fluid models}

We observe that  the difficulties mentioned above are removed if we deliberately eliminate the higher gradient terms in the first two equations of \eqref{final-D-eqn}. This step is usually accompanied by arguments resorting to well-established  WKB methods and corresponds to a classical treatment of the hydrodynamic flow: indeed,  one verifies that dropping the higher gradient terms in \eqref{final-D-eqn} takes the equation for the  phase function $S(\br,t)$ into a classical Hamilton-Jacobi equation. In the context of molecular dynamics, for example, this step corresponds to approximating nuclear motion as classical while leaving the electronic dynamics as fully quantum. 

In the  case of equations \eqref{final-D-eqn}, neglecting the second-order $\hbar^2$-terms  yields the  \emph{Ehrenfest fluid model} for the interaction of a classical system  and a quantum system: 
\beq
M(\partial_t + \bu\cdot\nabla)\bu   = 
-\langle\hat{\varrho}|\nabla\widehat{H}\rangle 
,\qquad\qquad
     i\hbar (\partial_t +\bu\cdot\nabla)\hat{\varrho} = [\widehat{H},\hat{\varrho}]
     ,\qquad\qquad
     \partial_t D+{\rm div}(D\bu)=0\,.
     \label{EhrenfestFluideqns}
\eeq
This complex fluid model is the simplest case of \emph{mixed quantum-classical dynamics}, which  generally comprises a vast set of models that have been devised to avoid the complications arising from fully quantum descriptions. While popular in molecular dynamics, similar  approaches have also been proposed in solid state physics \cite{BACaCaVe09,HuHeMa17}, spintronics \cite{Petrovic}, and quantum cosmology \cite{BoDi21}. 
Despite its  mathematical appeal, the Ehrenfest fluid model \eqref{EhrenfestFluideqns} is insufficient to provide a reliable description of the coupled dynamics \cite{AkLoPr14,TullyNonadiabaticDynamics}.  
Thus, much of the computational chemistry community has been developing alternative quantum-classical models beyond the Ehrenfest system \cite{CrBa18}.

Among the most popular alternatives to the Ehrenfest model, the \emph{surface hopping} algorithm was recently shown to violate Heisenberg's uncertainty principle, since the quantum density matrix $\hat\uprho$ may become sign-indefinite \cite{Bondarenko}. The same violation of the Heisenberg principle is also  allowed by the \emph{quantum-classical Liouville equation} on phase-space \cite{Aleksandrov,boucher,Gerasimenko,Kapral}.
In more generality, certain well-known consistency issues accompany  current quantum-classical approaches beyond the Ehrenfest equations \cite{AgCi07}.  On the one hand, since several studies do not seem to be  affected by these issues, the existing approaches continue to be regarded as a useful tool. On the other hand, the systematic formulation of  quantum-classical models beyond Ehrenfest dynamics   remains a subject of current research \cite{boucher,CrBa18,Marmo,Diosi,FoHoTr19,Hall,Kapral}.

While the  above models arise as hydrodynamic formulations of multi-particle systems, similar approaches have also appeared in different contexts to study systems involving a genuine complex fluid component. For example, mixed quantum-classical fluid models were proposed in solvation theory to describe the interaction dynamics of macroscopic  liquid solvents with quantum solute molecules \cite{Bousquet,Hughes}.
Since solvation phenomena occur  in a wide variety of natural and technological processes,   this solute-solvent interaction acquires crucial importance in different areas. In particular, the fluid-molecule coupling involves challenging nonlinear effects at different scales, which are revealed in unprecedented detail by time-resolved ultrafast experiments \cite{RoLaVa13}.
Current solvation  approaches  are mostly based on different combinations of computationally expensive atomistic approaches   with continuum descriptions involving many substantial approximations \cite{Santoro}. A direction that  remains little explored is the development of quantum-classical multiscale approaches combining microscopic solute dynamics with a mesoscopic solvent description. A remarkable effort in this direction is found in  \cite{Bousquet,BuBa06}, where the authors formulate a quantum-classical hydrodynamic model by applying the moment method from kinetic theory to the quantum-classical Liouville equation. Depending on the closure adopted for the operator-valued tensor-density $\widehat{\Pi}$, the system presented in \cite{Bousquet} reads
\beq
\partial_t\hat{\boldsymbol{g}}   = -\operatorname{div}\widehat{\Pi}
-\frac1{2}\big(\tilde{\rho}(\nabla\widehat{H}_\textit{eff})+(\nabla\widehat{H}_\textit{eff})\tilde{\rho}\big)+\frac{1}{i\hbar} \big[\hat{\boldsymbol{g}},\widehat{H}_\textit{eff}\big]
,\qquad\qquad
     i\hbar \partial_t \tilde{\rho} +\frac{ i\hbar }M \operatorname{div}\hat{\boldsymbol{g}} =  \big[\widehat{H}_\textit{eff},\tilde{\rho}\big].
\label{burghmod}     
\eeq
Here,  $\tilde{\rho}$ and $\hat{\boldsymbol{g}}$  are  operator-valued density and momentum-density, respectively, so that the fluid  density and momentum  are    $D=\operatorname{Tr}\tilde{\rho}$ and $\bm=\operatorname{Tr}\hat{\boldsymbol{g}}$. Also,  $\widehat{H}_\textit{eff}$ is an effective Hamiltonian carrying nonlocal terms. 
Benchmarked in \cite{Hughes} for certain closure schemes,
  the system \eqref{burghmod} provides a new promising perspective in solvation models.

Nonetheless, the necessity to address the  consistency requirements of quantum-classical dynamics motivates us  to consider also complementary modeling strategies to tackle problems  in both molecular dynamics, solvation theory, and other fields. In particular, this paper presents a new class of mixed quantum-classical models lying beyond Ehrenfest dynamics and still satisfying its consistency properties. Similarly to the work in \cite{Bousquet,BuBa06}, the present complex fluid model is formulated as a closure scheme of a more fundamental mixed quantum-classical approach on phase-space. However, instead of applying the closure at the level of the equations of motion, we will operate on the  variational principle underlying a recent phase-space formulation that, unlike other available  variants, appears to ensure a series of important consistency properties. As a result, the new hydrodynamic system inherits desirable variational and Hamiltonian structures which are analogous to those appearing in the theory of liquid crystals \cite{GBRa09,Holm02,Tronci12}.

\subsection{Quantum-classical dynamics on phase-space\label{sec:PSmod}}

Several mixed quantum-classical models currently used in molecular dynamics are based on the phase-space formulation of classical mechanics \cite{BuPa14,FaJiSp18,Kapral,WuEtAl}. While doubling the dimensionality of classical motion, this approach offers the opportunity for a wider range of possible approximations leading to computational schemes. Over the years, however, these phase-space formulations have not helped to overcome the consistency issues affecting quantum-classical models beyond Ehrenfest dynamics. Until  very recently.

Blending Koopman's Hilbert-space formulation of classical mechanics  with van Hove's unitary representations in symplectic geometry \cite{BoGBTr19,Koopman,VanHove}, the present authors  proposed a new phase-space model of quantum-classical mechanics that  satisfies all the following criteria \cite{boucher}  and still retains the geometric underpinnings of both classical and quantum mechanics \cite{BoGBTr19,GBTr22,GBTr21}:
1) the classical system is identified by a  probability density at all times;
2) the quantum system is identified by a positive-semidefinite density operator $\hat\uprho$ at all times;
3) the model is equivariant under both quantum unitary transformations and classical canonical transformations;
4)  in the absence of an interaction potential, the model reduces to uncoupled quantum and classical dynamics;
5) in the presence of an interaction potential, the {\it quantum purity} $\|\hat\uprho\|^2$ has nontrivial dynamics (decoherence property).

If $f(\bq,\bp,t)$ denotes the classical Liouville density on phase-space and $\uppsi(t;\bq,\bp) \in \mathsf{H}$  is the  conditional wavefunction representing the quantum dynamics (now depending on the classical phase-space coordinates), the model proposed in \cite{GBTr22,GBTr21} reads as follows 
\beq
{\partial_t f}
+\operatorname{div}(f\boldsymbol{\cal X})=0
\,,\qquad\qquad 
i\hbar({\partial_t \widehat{P}}
+\boldsymbol{\cal X}\cdot\nabla \widehat{P})=\big[\,\widehat{\!\mathscr{H}},\widehat{P}\big],
\label{HybEq1}
\eeq
with  $\widehat{P}=\uppsi\uppsi^\dagger$,
\beq
\boldsymbol{\cal X}=
\langle \widehat{P}|\bX_{\widehat{\cal H}}\rangle-\frac\hbar{2f}\Big( \big\langle (f\widehat{\boldsymbol\Sigma})\cdot\nabla|\bX_{{\widehat{\cal H}}}\big\rangle
-\big\langle\bX_{{\widehat{\cal H}}}\cdot\nabla| (f\widehat{\boldsymbol\Sigma})\big\rangle
\Big),
\qquad\quad
\widehat{\boldsymbol\Sigma}=i[\widehat{P},\bX_{\widehat{P}}]
\label{HybEq2}
\eeq
and
\beq
\,\widehat{\!\mathscr{H}}= \widehat{\cal H}+i\hbar\Big(\big\{\widehat{P},{\widehat{\cal H}}\big\}+\big\{\widehat{\cal H},\widehat{P}\big\}-\frac{1}{2f}\big[\big\{ f,\widehat{\cal H}\big\},\widehat{P}\big] \Big).
\label{HybEq3}
\eeq  
Here, $\{\cdot ,\cdot \}$ is the canonical Poisson bracket, $\widehat{\cal H}(\bq,\bp)$ is the Hermitian Hamiltonian operator  parameterized by the classical phase-space coordinates, and   we have denoted 
$
\bX_{\widehat{A}}=(\partial_{\bp}\widehat{A},-\partial_{\bq}\widehat{A})
$.
The density matrix of the quantum subsystem is expressed as $\hat\uprho=\int f \widehat{P}\,\de^3q\de^3p$.
If we discard the $\hbar$-corrections in both the vector field $\boldsymbol{\cal X}$ and the Hermitian generator $\,\widehat{\!\mathscr{H}}$, the model reduces to a phase-space variant \cite{ZiVa12} of the  Ehrenfest fluid equations \eqref{EhrenfestFluideqns}. Indeed,  after replacing $\boldsymbol{\cal X}=\langle \widehat{P}| \bX_{\widehat{\cal H}}\rangle$ and $\,\widehat{\!\mathscr{H}}={\widehat{\cal H}}$ in \eqref{HybEq1}, with ${\widehat{\cal H}}(\bq, \bp)=M^{-1}|\bp|^2/2+\widehat{H}(\bq)$, and  taking the hydrodynamic moments $(D,MD\bu)=\int (1,\bp)f\,\de^3p$ and $D\hat{ \varrho }=\int f\widehat{P}\,\de^3p$,  one recovers equations \eqref{EhrenfestFluideqns} by neglecting pressure effects.

While equations \eqref{HybEq1}-\eqref{HybEq3} appear hardly tractable, a direct calculation of $\operatorname{div}\!\boldsymbol{\cal X}$ reveals that no gradients of order higher than two appear in the equations \eqref{HybEq1}. In addition,  the variational and Hamiltonian structures of this system \cite{GBTr22,GBTr21} reveal the features occurring in quantum-classical coupling. For example, as we will see, equations \eqref{HybEq1}-\eqref{HybEq3} lead to the identification of a quantum-classical Poincar\'e integral invariant which extends the classical quantity $\oint \bp\cdot\de \bq$. In more generality, the system \eqref{HybEq1}-\eqref{HybEq3} appears to be the first quantum-classical Hamiltonian model beyond \makebox{Ehrenfest dynamics that satisfies the consistency criteria listed above.}

Given the combination of  consistency properties  and underlying mathematical structure, we consider the equations \eqref{HybEq1}-\eqref{HybEq3} as a platform for the formulation of simplified closure models that can be used in physically relevant cases. Indeed, the characteristic nature of the equations \eqref{HybEq1} offers the possibility of devising trajectory methods, e.g. following \cite{FoHoTr19,HoRaTr21,TrGB23LNCS}. For example, in \cite{BaBeTrGB24} we recently showed how particle schemes based on \eqref{HybEq1}-\eqref{HybEq3} succeed in reproducing crucial dynamical features in both quantum and classical sectors, well beyond the Ehrenfest model.
However, the model \eqref{HybEq1}-\eqref{HybEq3} involves the full classical phase-space, which makes the curse of dimensions particularly challenging. In an attempt to alleviate this problem, here we consider the formulation of a fluid closure on the configuration space. This  closure represents a quantum-classical alternative to the quantum hydrodynamic approach in \eqref{final-D-eqn} and yet retains correlation effects beyond Ehrenfest dynamics. 

\subsection{Plan of the paper} 

The paper proceeds by illustrating the original quantum-classical phase-space model in \S\ref{sec:PSmodbis}. After an overview of its derivation, the underlying action principle is presented in \S\ref{sec:actprin}, along with a discussion of its geometric structure. The Hamiltonian structure and its associated Casimir invariants are found in \S\ref{sec:HamStrInv}, while \S\ref{sec:hVNop} deals with the equivariance properties of the model.

The complex fluid model is formulated in \S\ref{sec:FluidCl}. After introducing the relevant fluid variables, the first closure steps are performed in \S\ref{sec:lifts} via a restriction to the configuration space of the original paths on phase-space. The  crucial  advance is presented in \S\ref{sec:backr}, where we show that the last term in \eqref{Tiziana3} may be projected to the configuration space by resorting to a Poisson bracket structure known as \emph{Nambu bracket}. This process leads to the appearance of two new scalar functions that are transported along the flow, one of which -- the \emph{backreaction field} -- incorporates the forces excerpted by the quantum system in the classical fluid flow. Then, the explicit form of the fluid equations is presented in \S\ref{sec:FluidEqns}, where we emphasize that the new model contains only first-order gradients, in contrast to the fully quantum treatment.

A discussion of the relevant properties of the new model is found in \S\ref{sec:MainFeat}. In \S\ref{sec:MeadConn}, we discuss the appearance of a non-Abelian gauge connection previously found in the context of molecular dynamics. In particular, this quantity seems to play an important role in the expression of the fluid stress tensor. The new complex fluid model possesses infinite families of dynamical invariants which are presented in \S\ref{sec:HamStrCrHel}, along with a discussion of the underlying Hamiltonian structure. Importantly, the Berry connection appears prominently when expressing the quantum dynamics in terms of wavefunctions, thereby allowing the characterization of a  \emph{cross-helicity} hydrodynamic invariant. An interesting specialization  is presented in \S\ref{sec:PureDef}, where we show that the proposed model succeeds in retaining backreaction forces otherwise lost in the Ehrenfest treatment. Finally, a class of invariant planar subsystems is illustrated in \S\ref{sec:planar}. In this case, we show that restricting to two spatial dimensions allows for larger families of dynamical invariants that may be used for Lyapunov stability studies.

\section{Geometric structure of the phase-space model\label{sec:PSmodbis}}

As anticipated, we will now review the formulation of the quantum-classical model \eqref{HybEq1}-\eqref{HybEq3}, with special emphasis on its geometric structure \cite{GBTr22}.
Before embarking on this discussion, we will start by presenting the theoretical background.
In first place, one adopts Koopman's Hilbert-space formulation of classical mechanics \cite{Koopman}: writing $f=|\chi|^2$ in the classical Liouville equation $\partial_t f=\{H,f\}$ yields $i\hbar\partial_t\chi=i\hbar\{H,\chi\}-\varphi\chi$, where $\varphi(\bq,\bp)$ is an arbitrary  function. Then, since the operator $i\hbar\{H,\_\}+\varphi\_$ is self-adjoint, the \emph{Koopman wavefunction} $\chi(\bq,\bp,t)$ evolves unitarily, just in the same way as  quantum Schr\"odinger wavefunctions. Guided by a series of arguments in representation theory,  van Hove fixed the particular phase term $\varphi=\bp\cdot\partial_\bp H-H$ \cite{VanHove}, that is  the phase-space form  of the Lagrangian  usually written as $\bp\cdot\dot{\bq}-H$. This step makes Koopman's original \makebox{formulation compatible with Hamilton-Jacobi theory \cite{GBTr20}.}

A first quantum-classical model was obtained by writing the \emph{Koopman-van Hove equation} $i\hbar\partial_t\chi=i\hbar\{H,\chi\}-(\bp\cdot\partial_\bp H-H)\chi$ for two interacting systems and then quantizing one of them \cite{BoGBTr19,GBTr20}. While ensuring a series of important consistency criteria, the resulting \emph{quantum-classical wave equation} $i\hbar\partial_t\Upsilon=i\hbar\{\widehat{\mathcal{H}},\Upsilon\}-(\bp\cdot\partial_\bp \widehat{\mathcal{H}}-\widehat{\mathcal{H}})\Upsilon$ generally allows for the classical Liouville density to become unsigned. Here,  $\Upsilon\in L^2(\Bbb{R}^6)\otimes{\sf H}$, where we recall that ${\sf H}$ is the quantum Hilbert space.
 The issue of classical positivity was addressed by resorting to a physical principle proposed by George Sudarshan \cite{Sudarshan}: any model of mixed quantum-classical dynamics should prevent classical phases from having observable effects. In other words, Hamilton-Jacobi functions are not measurable. This principle led the authors to treat the infinite-dimensional group of $S^1$-valued functions on phase-space as a gauge group. Indeed, as shown in \cite{GBTr21}, applying a $S^1$-symmetry by a suitable closure relation leads to treating classical phases as a \emph{gauge freedom}. This is in direct analogy with the role of unit complex numbers in standard quantum mechanics. While applying \makebox{the  $S^1$-symmetry  to the action principle}
\begin{equation}\label{action_principle}
\delta\int_{t_1}^{t_2}\!\int\operatorname{Re}\big\langle\Upsilon\big|i\hbar\partial_t\Upsilon-\{i\hbar\widehat{\mathcal{H}},\Upsilon\}+(\bp\cdot\partial_\bp\widehat{\mathcal{H}}-\widehat{\mathcal{H}})\Upsilon\big\rangle\,\de^3q\de^3p\,\de t=0
\end{equation}
for the quantum-classical wave equation was found to  restore positivity of the classical density, this procedure requires resorting to the Lagrangian  trajectories underlying classical phase-space motion. This requirement is accommodated by combining the exact factorization method \cite{abedi2012correlated} with the Euler-Poincar\'e variational theory of continuum media \cite{HoMaRa98}, which lies as the basis for the new treatment. However, the resulting model $i\hbar(\partial_t+\boldsymbol{\cal X}\cdot\nabla)\Upsilon+i\hbar(\operatorname{div}\boldsymbol{\cal X})\Upsilon/2=\,\widehat{\!\mathscr{H}}\Upsilon$ (up to  phase factors) is highly nonlinear and a real intuition on its structure can only come from investigating its variational formulation. Here, $\boldsymbol{\cal X}$ and $\,\widehat{\!\mathscr{H}}$ are given in \eqref{HybEq2}-\eqref{HybEq3} upon writing $f\widehat{P}=\Upsilon\Upsilon^\dagger$. Also, if we insist on considering only  finite-dimensional quantum systems, then $\Upsilon(\bq,\bp)$ is a Koopman wavefunction taking values in  $\mathsf{H}=\Bbb{C}^n$ and satisfying $\int \Upsilon^\dagger\Upsilon \de^6 z=1$. To avoid dealing with unnecessary phase factors, the remainder of this section considers this model as given in \eqref{HybEq1}-\eqref{HybEq3}, that is in terms of the quantities $f$ and $\widehat{P}$.

\subsection{Action principle and dynamics on group orbits\label{sec:actprin}} The basis for the model equations \eqref{HybEq1}-\eqref{HybEq3} is given by their underlying action principle $\delta\int_{t_1}^{t_2}\!L\,\de t=0$ associated  to the Euler-Poincar\'e Lagrangian
\beq
L(\boldsymbol{\cal X},f,\hat{\xi},\widehat{P})=\int \!f\big(\boldsymbol{\cal A}\cdot\boldsymbol{\cal X}+\big\langle { \widehat{P}},i\hbar\hat{\xi}-{\widehat{\cal H}}-i\hbar \{{ \widehat{P}},\widehat{\cal H}\}\big\rangle\big)\,\de^6z
\,.
\label{Tiziana3}
\eeq
We refer to \S5.1 in \cite{GBTr21} for the derivation of this Lagrangian from \eqref{action_principle} and the corresponding Euler-Poincar\'e variations.
Here, we have denoted $\bz=(\bq,\bp)$ and $\langle\,,\rangle=\operatorname{Re}\langle\,|\,\rangle$, while
$
{\boldsymbol{\cal A}=(\bp,0)}
$
 is the coordinate representation of the Liouville one-form ${\cal A}={\boldsymbol{\cal A}\cdot\de\bz}=\bp\cdot\de\bq$, so that $\omega=-\de {\cal A}$ is the canonical symplectic form. Also, $\hat{\xi}(\bz,t)$ is a skew-Hermitian operator and ${{ \widehat{P}}(\bz,t):=\uppsi(\bz,t)\uppsi(\bz,t)^\dagger}$. As  usual in Euler-Poincar\'e variational theories,  the variations $\delta f$ and $\delta \boldsymbol{\cal X}$ are constrained so that
\beq
\delta f=-\operatorname{div}(f\boldsymbol{\cal Y})
,\qquad\qquad\ 
\delta\boldsymbol{\cal X}=\partial_t\boldsymbol{\cal Y}+\boldsymbol{\cal X}\cdot\nabla\boldsymbol{\cal Y}-\boldsymbol{\cal Y}\cdot\nabla\boldsymbol{\cal X}
,
\eeq
where $\boldsymbol{\cal Y}$ is arbitrary. Analogously, we have
\beq
\delta { \widehat{P}} = [\hat\zeta, {  \widehat{P}}] -  \boldsymbol{\cal Y}\cdot\nabla \widehat{P}
,\qquad\qquad\ 
\delta \hat{\xi} = \partial _t {\hat\zeta} + [ {\hat\zeta} , \hat{\xi}  ] + \boldsymbol{\cal X}  \cdot\nabla  {\hat\zeta}   -  \boldsymbol{\cal Y}\cdot\nabla   \hat{\xi} 
,
\label{vars2}
\eeq
where $\hat\zeta$ is an arbitrary skew-Hermitian operator.
These variations arise by standard  Euler-Poincar\'e reduction from  Lagrangian to  Eulerian variables \cite{HoMaRa98}. In particular, if $\boldsymbol\eta(\bz_0,t)$ is the diffeomorphic Lagrangian path on phase-space and $U(\bz,t)$ is a unitary operator, we have the following relations between Lagrangian and Eulerian variables:
\beq
f=\boldsymbol\eta_*f_0
,\qquad\quad
{\widehat{P}}=(U{ \widehat{P}}_0U^\dagger)\circ\boldsymbol\eta^{-1}
,\qquad\quad
\dot{\boldsymbol\eta}=\boldsymbol{\cal X}\circ\boldsymbol\eta
,\qquad\quad
\hat{\xi}=\dot{U}U^\dagger\circ\boldsymbol\eta^{-1},
\label{LtoE}
\eeq
where $f_0$ and $\widehat{P}_0$ are  initial conditions. Here, ${\boldsymbol\eta_*f_0= (f_0/\det\nabla\boldsymbol\eta) \circ\boldsymbol\eta^{-1}}$ denotes the push-forward of the initial density by $\boldsymbol\eta$. We will denote the space of densities on the phase-space $T^*Q$ by $\operatorname{Den}(T^*Q)$, which is identified with the distributional dual ${\cal F}(T^*Q)^*$ of  scalar  functions ${\cal F}(T^*Q)$ via the $L^2$-pairing. Also, we have $\widehat{P}\in{\cal F}(T^*Q,\operatorname{Her}(\mathsf{H}))$, where $\operatorname{Her}(\mathsf{H})$ is the space of Hermitian operators on  $\sf H$ and ${\cal F}(T^*Q,\operatorname{Her}(\mathsf{H}))$ denotes the space of mappings $T^*Q\to\operatorname{Her}(\mathsf{H})$. While $f$ is positive-definite,  $\widehat{P}$  is only positive-semidefinite, and we will regard these   properties as being preserved in time from suitable initial conditions. 
The first two relations in \eqref{LtoE} indicate that the evolution of 
 \[
 (f,{ \widehat{P}})\in \operatorname{Den}(T^*Q)\times{\cal F}(T^*Q,\operatorname{Her}(\mathsf{H}))
 \]  occurs on orbits of the semidirect-product group 
 \[
 \operatorname{Diff}(T^*Q)\,\circledS\, {\cal F}(T^*Q,{\cal U}(\mathsf{H})),
 \] 
 where $\operatorname{Diff}(T^*Q)$ is the group of phase-space diffeomorphisms, and ${\cal F}(T^*Q,{\cal U}(\mathsf{H}))$ denotes the space of  functions on $T^*Q$ taking values into the group ${\cal U}(\mathsf{H})$ of unitary operators on the quantum Hilbert space $\mathsf{H}$. In particular, these orbits are determined by the group action given by the composition of the standard  conjugation representation of ${\cal F}(T^*Q,{\cal U}(\mathsf{H}))$ and the pushforward action of $\operatorname{Diff}(T^*Q)$. This  evolution on group orbits will be crucial in later sections. In this paper, we will consider $Q=\Bbb{R}^3$ for simplicity, although the treatment extends naturally to a  smooth manifold. 
 
This geometric structure reflects directly in the equations \eqref{HybEq1}-\eqref{HybEq3}. Indeed, taking the time derivative of the first two relations in \eqref{LtoE}  yields $\partial_t f+\operatorname{div}(f\boldsymbol{\cal X})=0$ and ${\partial_t{ \widehat{P}}+\boldsymbol{\cal X}\cdot\nabla{ \widehat{P}}=[\hat{\xi},{ \widehat{P}}]}$, respectively, and we  observe the appearance of Lie-transport operators. Furthermore, upon taking variations of \eqref{Tiziana3}, the action principle $\delta\int_{t_1}^{t_2}L\,\de t=0$ yields
\beq
\boldsymbol{{\cal X}}=\bX_{\textstyle\frac{\delta h}{
\delta f}}-\frac1f\bigg\langle{\frac{\delta h}{\delta { \widehat{P}}}}\Big| \bX_{\widehat{P}}\bigg\rangle
,\quad\ \ 
\bigg[i\hbar  \hat{\xi}-\frac1f\frac{\delta h}{
\delta { \widehat{P}}},{ \widehat{P}}\bigg]=0,
\quad\  \ \text{where}\quad
h=\!\int\!f\big\langle  {\widehat{\cal H}}+i\hbar \{{ \widehat{P}},\widehat{\cal H}\}\big\rangle\,\de^6z
\label{EPeqns}
\eeq
and we have used $\langle A\rangle:=\langle{ \widehat{P}},A\rangle$. Then, after various manipulations we recover the system \eqref{HybEq1}-\eqref{HybEq3}. The purely quantum and classical cases are found by restricting to the cases $\bX_{\widehat{\cal H}}=0$ and $\widehat{\cal H}={\cal H}\boldsymbol{1}$, respectively \cite{GBTr22}. We readily realize that it would be rather unwise  to try to obtain some information on this model by looking at the explicit equations \eqref{HybEq1}-\eqref{HybEq3} obtained by expanding the functional derivatives ${\delta h}/{\delta f}$ and ${\delta h}/{\delta { \widehat{P}}}$. Instead, we are motivated to accept the guidance of their underlying geometric structure. While we have seen how the latter manifests in the variational formulation \eqref{Tiziana3}-\eqref{vars2}, we  now  consider the corresponding Hamiltonian setting.

\subsection{Hamiltonian structure and Poincar\'e invariant\label{sec:HamStrInv}} 

As we will see below, the Hamiltonian structure of the model \eqref{HybEq1}-\eqref{HybEq3} is rather peculiar and we are not aware of similar structures occurring elsewhere in continuum mechanics.

First, it is convenient to introduce the weighted variable ${\cal \widehat{P}}=f \widehat{P}$ so that $f=\operatorname{Tr}{\cal \widehat{P}}$ and $\langle A\rangle=\langle {\cal \widehat{P}},A\rangle/\operatorname{Tr}{\cal \widehat{P}}$. Then,  the Euler-Poincar\'e Lagrangian \eqref{Tiziana3} may be expressed entirely in terms of the variables $(\boldsymbol{\cal X},\hat{\xi},{\cal \widehat{P}})$. Going through the same steps as above leads to rewriting \eqref{EPeqns} as
$
\boldsymbol{{\cal X}}=\langle \bX_{{\delta h}/{\delta {\cal \widehat{P}}}}\rangle
$ and $
{[i\hbar  \hat{\xi}-{\delta h}/{
\delta {\cal \widehat{P}}},{\cal \widehat{P}}]=0}
$, where ${
h=\int\big\langle  {\widehat{\cal H}}\operatorname{Tr}{\cal \widehat{P}}+i\hbar \{{\cal \widehat{P}},\widehat{\cal H}\}\big\rangle\,\de^6z}$. Then, the Hamiltonian equation 
\beq
i\hbar\frac{\partial\cal \widehat{P}}{\partial t}+i\hbar\operatorname{div}\!\bigg({\cal \widehat{P}}\Big\langle \bX_{\textstyle\frac{\delta h}{\delta {\cal \widehat{P}}}}\Big\rangle\bigg)=\bigg[\frac{\delta h}{\delta {\cal \widehat{P}}},{\cal \widehat{P}}\bigg]
\label{HamEqn}
\eeq
leads directly to the following bracket structure via the usual relation $\dot{f}=\{\!\!\{f,h\}\!\!\}$:
\begin{equation}\label{bracket_candidate_rho}
\{\!\!\{k,h\}\!\!\}(\mathcal{\widehat{P}})=\int \!\frac1{\operatorname{Tr}\mathcal{\widehat{P}}  }\bigg(\mathcal{\widehat{P}}  : \bigg\{\frac{\delta k}{\delta \mathcal{\widehat{P}}},\frac{\delta h}{\delta \mathcal{\widehat{P}}}\bigg\}: \mathcal{\widehat{P}}   \bigg)\de^6z 
 -
\int\!\left\langle \mathcal{\widehat{P}}  ,\frac{i}\hbar\!\left[\frac{\delta k}{\delta \mathcal{\widehat{P}}},\frac{\delta h}{\delta \mathcal{\widehat{P}}}\right] \right\rangle\de^6z,
\end{equation} 
where we have introduced the convenient notation ${A:B}=\operatorname{Tr}(AB)$. The proof that the bracket \eqref{bracket_candidate_rho} is Poisson involves a combination of results in Lagrangian and Poisson reduction  \cite{GBTr21}. On the one hand, we observe that the second term is naturally inherited from the usual Lie-Poisson structure underlying the quantum Liouville equation. On the other hand, the first term is a new type of Poisson bracket which reduces to the Lie-Poisson structure for the classical Liouville equation for $f=\operatorname{Tr}{\cal \widehat{P}}$ if both functional derivatives are a multiple of the identity.

Equation \eqref{HamEqn} easily leads to  characterizing the Casimir invariant $
C_1=\operatorname{Tr}\!\int\! f\Phi({\cal \widehat{P}}/f)\,\de^6z
$ for any matrix analytic function $\Phi$. Also, upon writing ${\cal \widehat{P}}=f\uppsi\uppsi^\dagger$, one finds the quantum-classical Poincar\'e integral invariant
$
\oint_{\boldsymbol{c}(t)}(\bp\cdot\de \bq+\langle\uppsi,i\hbar\de\uppsi\rangle)
=const.
$
for any loop $\boldsymbol{c}(t)=\boldsymbol\eta(\boldsymbol{c}_0,t)$ in phase-space, where we recall that $\boldsymbol\eta(t)$ is the flow of $\boldsymbol{\mathcal{X}}$. Then, by Stokes theorem, one identifies a Lie-transported quantum-classical two-form \makebox{on $T^*Q$, that is}
$
\Omega(t)=\boldsymbol\eta_*\Omega(0)
$
, 
with
$
\Omega(t):=\omega+\hbar\operatorname{Im}{\langle\de\uppsi(t)|\wedge\de\uppsi(t)\rangle}
$,
so that $\Omega(t)$ is symplectic if it is so initially. As a result, one finds the additional class of Casimir invariants
$
C_2=\int f\Theta\big(f^{-1}\Omega\wedge\Omega\wedge\Omega\big)\de^6z
$,
where $\Omega\wedge\Omega\wedge\Omega$  is a volume form  and $\Theta$ is  any scalar function of one variable. These Casimirs may be used to construct quantum-classical extensions of  Gibbs/von Neumann entropies \cite{GBTr22}.

\subsection{Quantum-classical von Neumann operator\label{sec:hVNop}} 
We observe that the Hamiltonian energy functional $h$ in \eqref{EPeqns} is not simply given by the usual average of the  Hamiltonian operator $\widehat{\cal H}$. Indeed, the $\hbar-$term seems to play a crucial role in taking the model \eqref{HybEq1}-\eqref{HybEq3} beyond simple Ehrenfest dynamics. As discussed in \cite{GBTr22}, this suggests that the quantum-classical correlations trigger extra energy terms that are not usually considered. A similar situation occurs in standard quantum mechanics with spin-orbit coupling, that is the coupling of the quantum spin  to the force acting on the orbital degrees of freedom. See Remark \ref{soc} below.  Obtained from the semirelativistic limit of the Dirac equation,  this effect is often discarded and yet it is crucial in a variety of contexts.

One may still rearrange the last relation in \eqref{EPeqns} in such a way that 
the total energy is formally given by an average of $\widehat{\cal H}$. Following this route leads to rewriting the last relation in \eqref{EPeqns} 
 as $h=\operatorname{Tr}\int\!\widehat{\cal D}\widehat{\cal H}\,\de^6z$, where 
\beq\label{hVNop}
\widehat{\cal D}
=
f{\widehat{P}}+\frac{i\hbar}2\operatorname{div}\!\big(f[\widehat{P},\bX_{\widehat{P}}]\big)
\eeq
is a measure-valued von Neumann operator. Then, classical and quantum densities are simply given  by taking the   trace and integral of $\widehat{\cal D}$, respectively. Unlike the quantum density operator, we observe that the hybrid operator $\widehat{\cal D}$ is not sign-definite. Remarkably, however, $\widehat{\cal D}$ enjoys the  equivariance properties
\beq
\widehat{\cal D}(\tilde{\boldsymbol\eta}_*f,\tilde{\boldsymbol\eta}_*\widehat{P})=\tilde{\boldsymbol\eta}_*\widehat{\cal D}(f,\widehat{P}) 
\qquad\text{ and }\qquad
\widehat{\cal D}(f,\mathscr{U}\widehat{P}\mathscr{U}^\dagger)=\mathscr{U}\widehat{\cal D}(f,\widehat{P})\mathscr{U}^\dagger\,,
\label{equiv}
\eeq
where $ \tilde{\boldsymbol\eta}$ is a symplectic diffeomorphism on $T^*Q$ and $\mathscr{U}\in{\cal U}(\mathsf{H})$.
These two properties ensure the following dynamics in the classical and quantum sector, respectively \cite{GBTr21}:
$
{\partial_t f}=\operatorname{Tr}\{\widehat{\cal H},\widehat{\cal D}\}
$ and $
i\hbar{\de\hat\uprho}/{\de t}=\int[\widehat{\cal H},\widehat{\cal D}]\,\de^6z$. 
For example, upon denoting $\widehat{\boldsymbol\Sigma}=i[\widehat{P},\bX_{\widehat{P}}]$, the first property is verified directly by writing 
\[
{\boldsymbol{\cal X}} =f^{-1}\langle\widehat{\cal D}, \bX_ {\widehat{\cal H}}\rangle+\hbar f^{-1}\operatorname{div}\!\big(f\operatorname{Tr}( {\bX_ {\widehat{\cal H}}\wedge\widehat{\boldsymbol\Sigma}})\big)
,\quad\text{ where }\quad
( \bX_ {\widehat{\cal H}}\wedge\widehat{\boldsymbol\Sigma})^{jk}:=\frac12\big( X_ {\widehat{\cal H}\,}^j{\widehat{\Sigma}}^k-{\widehat{\Sigma}}^j X_ {\widehat{\cal H}}^k\big)
\]
identifies a bivector. With this expression of $\boldsymbol{\cal X}$ one recognizes  that $\operatorname{div}\!\boldsymbol{\cal X}=\operatorname{Tr}\{\widehat{P}+\hbar\operatorname{div}\widehat{\boldsymbol\Sigma},\widehat{\cal H}\}/2$ $-\hbar\operatorname{div}\operatorname{Tr}(\{\log f,\widehat{\cal H}\}\widehat{\boldsymbol\Sigma})/2$ involves only first- and second-order gradients. In more generality, due to the equivariance properties \eqref{equiv}, the hybrid von Neumann operator acquires an important role in different aspects, such as the quantum-classical momentum balance  \cite{GBTr22}.

\section{Formulation of the fluid closure\label{sec:FluidCl}}
\rem{ 
Given its phase-space formulation, the  quantum-classical model \eqref{HybEq1}-\eqref{HybEq3} appears quite complex. Nevertheless, some conclusions may be drawn, especially in comparison to previous models. Indeed, upon considering quantum two-level systems, a simple analysis of {\it pure-dephasing dynamics} \cite{} shows that the present model overcomes the difficulties occurring in the Ehrenfest equations \eqref{EhrenfestFluideqns}. In this case, the quantum-classical Hamiltonian operator reads $\widehat{\cal H}={\cal H}_0(\bz)+{\cal H}_I(\bz)\widehat{\sigma}_k$, where ${\cal H}_0$ and ${\cal H}_I$ are scalar functions, and $\widehat{\sigma}_k$ denotes the $k-$th Pauli matrix. Then, one observes that the initial condition $\langle\widehat{\sigma}_k\rangle=0$ is preserved in time by the second equation in \eqref{EhrenfestFluideqns} so that the classical  dynamics given by the first equation therein decouples completely from the quantum evolution. On the contrary, no such decoupling takes place for the quantum-classical model \eqref{HybEq1}-\eqref{HybEq3}, as long as the $\hbar-$correction terms are retained \cite{}. Pure-dephasing dynamics, however, comprises a very narrow class of possible study cases and more general problems need to be considered. 
} 

Given the level of complexity of the model \eqref{HybEq1}-\eqref{HybEq3}, it is desirable to develop suitable closure schemes that can substantially simplify the treatment.  In particular, here we present a fluid closure  in which the classical system is represented by a fluid with density and momentum
\beq
{D}(\bq)=\int\! f(\bq,\bp)\,\de^3 p
,\qquad\qquad
\bm(\bq)=\int\! \bp f(\bq,\bp)\,\de^3 p
\,,
\label{moments1}
\eeq
respectively, 
while the measure-valued density ${\cal \widehat{P}}(\bq,\bp)$ on phase-space is replaced by 
\beq
\int \!f(\bq,\bp)\widehat{P}(\bq,\bp)\,\de^3 p=:{D}(\bq){\hat{\varrho}}(\bq)
\,,
\label{moments2}
\eeq
where ${\operatorname{Tr}\hat\varrho(\bq)=1}$.
In particular, we focus on quantum-classical Hamiltonians of the type
\beq\label{hamilt}
\widehat{\cal H}(\bq,\bp)=\frac1{2M}|\bp|^2+\widehat{H}(\bq).
\eeq
While we could implement the moment method directly on the equations of motion, as customary in the kinetic theory of gases, here we perform the fluid closure at the  level of the variational structure \eqref{Tiziana3}-\eqref{vars2} underlying the full phase-space model. Before proceeding, we emphasize that the current treatment does not include dissipation and diffusion. Indeed, we rely on the possibility to add standard diffusion and dissipation terms \emph{a posteriori}, once the conservative equations are written explicitly. Similarly,  pressure effects arising from the purely classical kinematic term $\bp\cdot\partial_{\bq} f/M$ in the first  equation of \eqref{HybEq1} will also be added \emph{a posteriori}  by resorting to a conventional equation of state.

\subsection{Restriction to cotangent lifts\label{sec:lifts}}

So far the classical  motion has been identified with diffeomorphic paths $\boldsymbol\eta$ on phase-space via their generating vector field $\boldsymbol{\cal X}$ defined in \eqref{LtoE}. Each of these paths acts on the classical density as well as on the unitary operators governing the quantum dynamics as in \eqref{LtoE}. In order to obtain a fluid description of the classical motion, the latter needs to be restricted to occur on the configuration space. Since we want to operate on the Eulerian action principle associated to the Lagrangian \eqref{Tiziana3}, we need to restrict the phase-space vector field $\boldsymbol{\cal X}$ to generate only transformations on the configuration space $Q$.  We also ask for these transformations to possess a natural extension to phase-space. Such a class of transformations is well known under the name of \emph{cotangent lifts} (or simply \emph{lifts}) of diffeomorphisms. In particular, any diffeomorphism ${\boldsymbol\upeta}\in \operatorname{Diff}(Q)$ induces a phase-space diffeomorphism ${\boldsymbol\upeta}^{\sf L}\in \operatorname{Diff}(T^*Q)$, that is the `lift' of ${\boldsymbol\upeta}$, which reads
${\boldsymbol\upeta}^{\sf L}(\bq,\bp)=\big({\boldsymbol\upeta}(\bq),\nabla{\boldsymbol\upeta} ^{-1} (\boldsymbol\upeta(\bq))\cdot\bp\big)$. Notice that we have $\boldsymbol\upeta^{\sf L}_*{\cal A}={\cal A}$, where ${\cal A}=\bp\cdot\de\bq$. Likewise, in terms of Eulerian variables, any vector field $\bu(\bq)$ identifies a  lift on phase-space, which reads
$
\bu^{\sf L}(\bq,\bp)=\big(\bu(\bq),-{\nabla\bu(\bq)\cdot\bp}\big)
$.  Thus, as a first step, we will restrict phase-space paths to  cotangent lifts of paths on the configuration space, and this amounts to replacing
\[
\int \!f(\bq,\bp) \boldsymbol{\cal A}(\bq,\bp)\cdot\boldsymbol{\cal X}(\bq,\bp)\,\de^6z \  \longrightarrow\, \int \bm(\bq)\cdot\bu(\bq)\,\de^3q
\]
in \eqref{Tiziana3}. Here, we have used the second definition in \eqref{moments1} while the vector field $\bu$ is defined in terms of Lagrangian paths as $\dot{{\boldsymbol\upeta}}=\bu\circ{\boldsymbol\upeta}$. Notice that, on phase space, one also has $\dot{{\boldsymbol\upeta}}^{\sf L}=\bu^{\sf L}\circ{\boldsymbol\upeta}^{\sf L}$.

In pursuing our restriction to subgroup transformations on the configuration space, we also need to deal with the quantum propagators $U\in {\cal F}(T^*Q,{\cal U}(\mathsf{H}))$, which are again mappings defined on the classical phase space. In order to restrict to the configuration space, here we will consider only transformations ${\sf U}(\bq)$ in the subgroup ${\cal F}(Q,{\cal U}(\mathsf{H}))$. Then, the quantum generator of motion $\hat{\xi}$ in \eqref{LtoE} is replaced by $\hat{\upxi}=\dot{\sf U}{\sf U}^{-1}\circ{\boldsymbol\upeta}^{-1}$, with ${\sf U}\in {\cal F}(Q,{\cal U}(\mathsf{H}))$ and ${\boldsymbol\upeta}\in \operatorname{Diff}(Q)$. Consequently, by recalling \eqref{moments2}, we make the following replacement in \eqref{Tiziana3}:
\[
\int \!f(\bq,\bp)\big\langle { \widehat{P}(\bq,\bp)},i\hbar\hat{\xi}(\bq,\bp)\big\rangle\,\de^6z
\ \longrightarrow\,\int \!D(\bq)\big\langle { \hat{\varrho}(\bq)},i\hbar\hat{\upxi}(\bq)\big\rangle\,\de^3q.
\]

Another important step concerns the term $\int \!f\langle { \widehat{P}},\widehat{\cal H}\rangle\,\de^6z$ in \eqref{Tiziana3}. At this stage we adopt the \emph{cold-fluid} closure $f( \bq,\bp)=D( \bq)\delta(\bp-\bm(\bq)/D(\bq))$, thereby discarding pressure effects. Upon using \eqref{hamilt},  this expression of the phase-space density leads to the further replacement
\[
\int \!f\langle  \widehat{P},\widehat{\cal H}\rangle\,\de^6z
\ \longrightarrow\,
\int \Big(\frac{|\bm|^2}{2MD}+D\langle { \hat{\varrho}},\widehat{H}\rangle\Big)\de^3q.
\]
Notice that the cold-fluid closure is  used here only as an intermediate step restricting to the simplest possible case. We will restore pressure effects later on by adding a fluid internal energy term to the  right of the  relation above.

\subsection{The quantum backreaction\label{sec:backr}}
So far, the quantities in the Lagrangian \eqref{Tiziana3} were restricted to the configuration space, except for the last term. The latter is responsible for the \emph{quantum backreaction} on the classical flow beyond the usual Hellman-Feynman force averages \cite{Feynman2} already appearing on the right-hand side of the momentum  equation in the Ehrenfest model \eqref{EhrenfestFluideqns}. The last term in \eqref{Tiziana3} is responsible for a much intricate structure of the equations \eqref{HybEq1}-\eqref{HybEq3}, which involve  gradients of the quantum variable $\widehat{P}$ and make the identification of a closure far from obvious. Here, we will proceed in stages and pursue the simplest possible closure method to retain nontrivial backreaction effects. Similarly to the procedure  in the previous section, here we will continue to operate on the Lagrangian \eqref{Tiziana3}, thereby avoiding the necessity to control the various gradients  in  \eqref{HybEq1}-\eqref{HybEq3}.

We start by rearranging the last term in \eqref{Tiziana3} as
\begin{align}\nonumber
\int \!f\big\langle { \widehat{P}},i\hbar \{{ \widehat{P}},\widehat{\cal H}\}\big\rangle\,\de^6z
=&\ \frac12\int \!\big\langle {\widehat{\cal H}},i\hbar \operatorname{div}[{ \widehat{P}},f\bX_{ \widehat{P}}]\big\rangle\,\de^6z
\\
=&\ 
\frac12\int \!\bigg\langle {\widehat{H}},i\hbar\operatorname{div}\!\int [\widehat{P},f\partial_\bp{ \widehat{P}}]\,\de^3 p\bigg\rangle\,\de^3q\,,
\label{auxiliary}
\end{align}
where the second divergence  involves only position coordinates in configuration space.
The first equality above follows from  a slight rearrangement involving the projection of $i \{{ \widehat{P}},\widehat{\cal H}\}=i\partial_k \widehat{P}X_{\widehat{\cal H}}^k$ on its Hermitian part as well as integration by parts. Also, the second equality  follows from \eqref{hamilt}. Upon resorting to the cold-fluid closure, we recall  \eqref{moments2} to write $ \hat{\varrho}=\hat P|_{\bp=\bm/D}$, and eventually perform the replacement
\[
\operatorname{div}\!\int [\widehat{P},f\partial_\bp{ \widehat{P}}]\,\de^3 p
\ \longrightarrow\ 
 \operatorname{div}[\hat{\varrho},\widehat{\boldsymbol\kappa}]
 \,,\qquad\text{ with }\qquad
\widehat{\boldsymbol\kappa}= D\partial_\bp{ \widehat{P}}|_{\bp=\bm/D}.
\]
At this point, our closure problem amounts to finding a closure for the operator-valued current $\widehat{\boldsymbol\kappa}$. In particular, we  need to find a suitable expression of $\widehat{\boldsymbol\kappa}$ in terms of $\hat{\varrho}$ and $\nabla \hat{\varrho}$. For the sake of simplicity, here we  assume a linear relation between $\widehat{\boldsymbol\kappa}$ and $\nabla \hat{\varrho}$. 

To proceed further,  we will be guided by the properties of the original model \eqref{HybEq1}-\eqref{HybEq3}. For example, we notice that the term $\operatorname{div} [\widehat{P},f\bX_{ \widehat{P}}]= f[\partial_k \widehat{P},X^k_{ \widehat{P}}]+[\widehat{P},X^k_{ \widehat{P}}]\partial_kf$ in the first line of \eqref{auxiliary}
does not produce gradients of order higher than one. Then, in order to avoid the emergence of higher-order gradients in  $ \operatorname{div}[\hat{\varrho},\widehat{\boldsymbol\kappa}]= [\partial_k\hat{\varrho},\widehat{\kappa}^k]+ [\hat{\varrho},\operatorname{div}\widehat{\boldsymbol\kappa}]$, the assumption of linearity  between $\widehat{\boldsymbol\kappa}$ and $\nabla \hat{\varrho}$ leaves us with the choice
\[
\widehat{\boldsymbol\kappa}=-\boldsymbol{\beta}\times\nabla\hat{\varrho}
\,,
\]
where $\boldsymbol{\beta}(\bq,t)\cdot\de \bq$ is a  real-valued  differential one-form
and the minus sign is chosen for later convenience. Notice that  ${\operatorname{Tr}\hat{\varrho}=1\implies{\operatorname{Tr}\widehat{\boldsymbol\kappa}=0}}$ and, in the case ${\hat\varrho=\psi\psi^\dagger}$,  we also have $\langle\hat{\varrho}|\widehat{\boldsymbol\kappa}\rangle=0$. 
While this possible closure will be explored elsewhere, here we deal with the  case  $\boldsymbol{\beta}=c\nabla b$ for some scalar functions $b(\bq,t)$ and $c(\bq,t)$, that is
$\widehat{\boldsymbol\kappa}=c\nabla \hat{\varrho}\times\nabla b$.

With this expression of $\widehat{\boldsymbol\kappa}$,   the Lagrangian \eqref{Tiziana3} finally reduces to
\beq
l(\bm,\boldsymbol{u},D,\hat{\upxi},\hat{\varrho},b,c)=\int \!\Big(\bm\cdot\boldsymbol{u}-\frac{|\bm|^2}{2MD}+\big\langle \hat{\varrho},i\hbar D\hat{\upxi}-D{\widehat{H}}-i\hbar c\{{ \hat{\varrho}},\widehat{ H}\}_b\big\rangle\Big)\de^3q
\,,
\label{Tiziana4}
\eeq
where
\beq
\{F,G\}_b:=\nabla b\cdot\nabla F\times\nabla G
\label{nambubkt}
\eeq
defines a Poisson bracket structure of Nambu type \cite{Nambu} on any functions $F$  and $G$  and for any  function $b$. 
Since the latter is responsible for the backreaction terms in the new Lagrangian \eqref{Tiziana4}, we will refer to $b$ as the \emph{backreaction field}. We observe the striking analogy between the last term in \eqref{Tiziana4} and the last term in \eqref{Tiziana3}.   For later purpose, we notice that $\langle\hat{\varrho},i\hbar \{{ \hat{\varrho}},\widehat{ H}\}_b\rangle=\langle\hat{\varrho},i\hbar [\nabla{ \hat{\varrho}},\nabla b\times\nabla\widehat{ H}]\rangle/2$. 
\begin{remark}[Extension to higher dimensions]
The present treatment can be extended to consider the entire coordinate space $\Bbb{R}^{3d}$ for a molecule with $d$ nuclei. In this case, each term in the Lagrangian \eqref{Tiziana4} would be replaced by its natural higher-dimensional generalization. In particular, the last term would now involve the the direct sum of the same Nambu bracket $d$ times. More explicitly, the Nambu bracket \eqref{nambubkt} would be replaced by the more general Poisson bracket $\{F,G\}_b(\br_1,\dots,\br_d)=\sum_{k=1}^d\nabla_{\!\br_k} b\cdot\nabla_{\!\br_k} F\times\nabla_{\!\br_k} G$. 
\end{remark}
Besides $b$ and $c$, which are dealt with below, the Lagrangian \eqref{Tiziana4} involves the following quantities:
\beq
D:=\upeta_*D_0
,\qquad\quad
{\hat{\varrho}}:=({\sf U}{ \hat{\varrho}}_0{\sf U}^\dagger) \circ \boldsymbol\upeta ^{-1} 
,\qquad\quad
\boldsymbol{u}:= \dot{\boldsymbol\upeta}\circ\boldsymbol\upeta ^{-1} 
,\qquad\quad
\hat{\upxi}:=\dot{\sf U}{\sf U}^\dagger\circ\boldsymbol\upeta^{-1}.
\label{LtoE2}
\eeq
Here, the last two expressions arise from the restrictions performed in \S\ref{sec:lifts}, while the first two are obtained by  combining the definitions in \eqref{moments1}-\eqref{moments2} with the first two relations in \eqref{LtoE}, as well as replacing $\boldsymbol\eta\to\boldsymbol\upeta$ and $\hat{\xi}\to\hat{\upxi}$. In particular, the first two in \eqref{LtoE2} yield the auxiliary equations
\beq\label{auxeqns}
\partial_t D+\operatorname{div}(D\boldsymbol{u})=0
,\qquad\qquad
\partial_t {\hat{\varrho}}+\boldsymbol{u}\cdot{\hat{\varrho}}=[\hat{\upxi},{\hat{\varrho}}],
\eeq
which accompany the variational principle associated to \eqref{Tiziana4}.

The scalar field $c$ is inserted in the expression of $\widehat{\boldsymbol\kappa}$ in such a way that the last integral in \eqref{Tiziana4} can be made to converge. In particular, $c(\bq,t)$ is conveniently prescribed as the ratio between the classical density $D(\bq,t)$ and the Lie-transported  volume element $\mu=\upeta_*\mu_0$, where $\mu_0$ is the elementary volume form in physical space. The appearance of the volume form in the denominator arises from the formal definition of Nambu-Poisson structures, as found in \cite{Vaisman}. Then, upon setting $\mu_0=1$ in $\Bbb{R}^3$, we have
\[
c=D_0\circ\boldsymbol\upeta ^{-1}
.
\]
Here, we emphasize the difference between the push-forward of the classical density  $\upeta^* D_0=(D_0/\operatorname{det}\nabla\boldsymbol\upeta)\circ\boldsymbol\upeta ^{-1}$ and the simpler composition operation $D_0\circ\boldsymbol\upeta ^{-1}$, corresponding to the push-forward of the scalar function $D_0/\mu_0$. The equation of motion for $c(\bq,t)$ is thus  $\partial_t c+\bu\cdot\nabla c=0$.

So far, nothing has been said about the evolution of the backreaction field $b$ in \eqref{nambubkt}. The simplest choice would  involve a time-independent function. However, we will now show that this is not compatible with the structure of the original model \eqref{HybEq1}-\eqref{HybEq3}. A certain insight is provided by the observation that the terms involving $\widehat{H}$ in \eqref{Tiziana4} combine into $\int \langle \widehat{\mathscr{D}},\widehat{H}\rangle\,\de^3q$, where
\beq
 \widehat{\mathscr{D}}=D\hat{\varrho}+\frac{i\hbar}2 \operatorname{div}(c[\hat{\varrho},\nabla \hat{\varrho}\times\nabla b])
 \label{vnop}
\eeq
 evidently plays the same role as the hybrid von Neumann operator $\widehat{\cal D}$ in \eqref{hVNop}. By pursuing the analogy with the original model \eqref{HybEq1}-\eqref{HybEq3}, here we will insist that $\widehat{\mathscr{D}}$ must satisfy analogous  equivariance properties to those in \eqref{equiv}. In the case that $b$ is fixed and $\widehat{\mathscr{D}}$ is regarded as a map of the dynamical variables $(D,\hat{\varrho},c)$, evidently we would have, in push-forward notation, $\widehat{\mathscr{D}}(\upeta_*D,\upeta_*\hat{\varrho},\upeta_*c)\neq\upeta_*\widehat{\mathscr{D}}(D,\hat{\varrho},c)$. Instead, if $b$ is time-dependent, so that $\widehat{\mathscr{D}}$ depends on the four variables $(D,\hat{\varrho},b,c)$, then we observe that
\beq
\widehat{\mathscr{D}}(\upeta_*D,\upeta_*\hat{\varrho},\upeta_*b,\upeta_*c)=\upeta_*\widehat{\mathscr{D}}(D,\hat{\varrho},b,c) 
\,,\qquad \qquad\!\!
\widehat{\mathscr{D}}(D,\mathscr{U}\hat{\varrho}\mathscr{U}^\dagger,b,c)=\mathscr{U}\widehat{\mathscr{D}}(D,\hat{\varrho},b,c)\mathscr{U}^\dagger
\label{equiv2}.
\eeq 
In particular, here we make the following prescription for the evolution of $b$:
\beq
b=b_0\circ{\boldsymbol\upeta}^{-1}
\,,
\label{bevol}
\eeq
that is $b$ is also Lie transported as a scalar function. As we will see, this relation ensures conservation of the total momentum exchanged between the classical and the quantum systems.

\subsection{Quantum-classical fluid equations\label{sec:FluidEqns}}
Having characterized the fluid Lagrangian \eqref{Tiziana4} along with the evolution of the dynamical variables, we are now ready to write the equations of motion explicitly. In particular, Hamilton's variational principle $\delta\int_{t_1}^{t_2}l\,\de t=0$ involves the following Euler-Poincar\'e variations:
\beq
\delta D=-\operatorname{div}(D\bw)
,\qquad\qquad\ 
\delta\bu=\partial_t\bw+\bu\cdot\nabla\bw-\bw\cdot\nabla\bu
, \qquad\qquad\ \delta c = - \bw \cdot \nabla c,
\label{vars2_5}
\eeq
and
\beq
\delta { \hat\varrho} = [{\hat\upzeta}, {  \hat\varrho}] -  \bw\cdot\nabla \hat\varrho
,\qquad 
\delta \hat{\upxi} = \partial _t {\hat\upzeta} + [ {\hat\upzeta} , \hat{\upxi}  ] + \bu  \cdot\nabla  {\hat\upzeta}   -  \bw\cdot\nabla   \hat{\upxi}, \qquad \delta b = - \bw \cdot \nabla b,
\label{vars3}
\eeq
where $\bw=\delta\boldsymbol\upeta\circ\boldsymbol\upeta^{-1}$ and $\hat\upzeta=\delta{\sf U}{\sf U}^{-1}\circ\boldsymbol\upeta^{-1}$ are both arbitrary and vanishing at the endpoints. These variations follow directly from \eqref{LtoE2} and \eqref{bevol}. In addition, $\delta\bm$ is arbitrary.

Before proceeding further, in this section we will make use of the partial Legendre transform $\bm=DM\bu$ in \eqref{Tiziana4} and we will restore pressure by inserting an internal energy term of the type $-\int D{\cal E}(D)\,\de^3 q$. While pressure effects are absent in   molecular dynamics, they become important for solute-solvent coupling in solvation hydrodynamics \cite{Bousquet,Hughes}. In that context, the  motion of a quantum molecule interacting with the surrounding fluid solvent cannot generally ignore the solvent pressure. Upon including the internal energy, the Lagrangian \eqref{Tiziana4} becomes
\beq
\ell(\boldsymbol{u},D,\hat{\upxi},\hat{\varrho},b,c)=\int \!\Big(\frac{M}{2}D|\bu|^2-D{\cal E}(D)+\big\langle \hat{\varrho},i\hbar D\hat{\upxi}-D{\widehat{H}}-i\hbar c\{{ \hat{\varrho}},\widehat{ H}\}_b\big\rangle\Big)\de^3q
\,.
\label{Tiziana5}
\eeq
In analogy with the phase-space construction in \S\ref{sec:actprin}, and as a result of the relations \eqref{LtoE2} and \eqref{bevol}, the action principle associated to this Lagrangian restricts the evolution of $(D,\hat\varrho,b, c )$ to occur on orbits of the semidirect product $
 \operatorname{Diff}(Q)\,\circledS\, {\cal F}(Q,{\cal U}(\mathsf{H}))$ acting on the space $\operatorname{Den}(Q)\times{\cal F}(Q,\operatorname{Her}(\mathsf{H}))\times {\cal F}(Q,\Bbb{R}^2)\times {\cal F}(Q,\Bbb{R}^2)$. We notice the presence of the extra variables $b$ and $c$ arising from the closure method and previously absent in the phase-space approach. In addition, in the present case, the Lagrangian fluid trajectories  evolve along a velocity vector field $\bu$ that has its own equation of motion and will characterize the fluid momentum evolution. This situation differs from that in the phase-space model, where the Lagrangian paths were determined by a vector field  which is given in the first equation of  \eqref{EPeqns} by an algebraic expression involving the other dynamical variables. Despite these differences, the fluid system resulting from the action principle $\delta\int_{t_1}^{t_2}\ell\,\de t=0$ has a geometric structure that is shared with common models of complex fluids \cite{GBRa09,Holm02,Tronci12}. 
 
While further geometric structures appearing in the present closure model will be presented  later on,  at the moment we are interested in the explicit form of the equations of motion. Upon denoting ${\sf p}=D^2{\cal E}'(D)$, the action principle associated to the Lagrangian \eqref{Tiziana5}  leads to
\begin{equation}\label{QCeqs1}
\begin{aligned}
MD(\partial_t+\bu\cdot\nabla)\bu=& -\nabla{\sf p}
-D\langle\hat\varrho,\nabla\widehat{H}\rangle
+\hbar \big\langle\nabla(c\hat\varrho),i\{\hat\varrho,\widehat{H}\}_b\big\rangle
+\hbar \big\langle\nabla\hat\varrho,{i} \{c\hat\varrho, \widehat{H}\}_b\big\rangle\\
&
-\hbar \operatorname{div}\langle c\hat\varrho,i \nabla\hat\varrho\times\nabla\widehat{H}\rangle\nabla b,\\
i\hbar (\partial_t+\bu\cdot\nabla)\hat\varrho=&\, \Big[\widehat{H}+{i\hbar}D^{-1}\Big(c\{\hat\varrho,\widehat{H}\}_b+c\{\widehat{H},\hat\varrho\}_b-\frac{1}{2}\big[\{c, \widehat{H}\}_b,\hat\varrho\big]\Big),\hat\varrho\Big],
\end{aligned}
\end{equation}
as well as
\beq
\partial_t D+\operatorname{div}(D\bu)=0
,\qquad\qquad\ 
(\partial_t+\bu\cdot\nabla)b=0
,\qquad\qquad\ 
(\partial_t+\bu\cdot\nabla)c=0\,.
\label{QCeqs2}
\eeq
We refer to Appendix \ref{appendixA} for more calculational details.  These quantum-classical fluid equations can be immediately compared to the fully quantum equations in \eqref{final-D-eqn}. We briefly summarize our considerations as follows:
\begin{itemize}

\item with respect to the fully quantum treatment in \eqref{final-D-eqn}, the quantum-classical system \eqref{QCeqs1}-\eqref{QCeqs2} involves two extra variables, the backreaction field $b$ and the quantity $c$, which are Lie transported as scalar functions;

\item the quantum potential $V_Q$ is now absent and its effects are now replaced by classical  pressure effects from standard barotropic fluids;

\item unlike the quantum equations \eqref{final-D-eqn}, the quantum-classical system \eqref{QCeqs1}-\eqref{QCeqs2} recovers uncoupled quantum and classical dynamics if $\nabla\widehat{H}=\nabla H_C\boldsymbol{1}$, for some scalar function $H_C$;

\item while the fully quantum equations \eqref{final-D-eqn} contain terms of order $\hbar^2$, only terms of the order $\hbar$ appear in the the quantum-classical system   \eqref{QCeqs1}-\eqref{QCeqs2};

\item unlike the quantum equations \eqref{final-D-eqn}, the quantum-classical system \eqref{QCeqs1}-\eqref{QCeqs2} involves {\it only first-order gradients};

\item unlike the solvation model  \eqref{burghmod}, equations \eqref{QCeqs1}-\eqref{QCeqs2} ensure that the operator $\hat\varrho$ is positive semi-definite at all times. This follows from the second equation in \eqref{auxeqns};

\item the system \eqref{QCeqs1}-\eqref{QCeqs2} may be readily extended to include a nonlocal dependence of the Hamiltonian $\widehat{H}$ on the variables $D$ and $\tilde\rho=D\hat\varrho$, as occurring in the effective Hamiltonian $\widehat{H}_\textit{eff}$ of the solvation model \eqref{burghmod} in \cite{Bousquet}.

\end{itemize}
Thus, despite their cumbersome appearance and the presence of two auxiliary frozen-in functions, the quantum-classical fluid equations represent a considerable simplification over the fully quantum hydrodynamics, where second- and third-order gradients are responsible for the appearance of interference patterns that pose severe challenges for current trajectory-based algorithms \cite{ZhMa03}. 

We also notice that, upon dropping   the internal energy and setting an initially constant backreaction field,  the system \eqref{QCeqs1} recovers  the quantum-classical Ehrenfest equations \eqref{EhrenfestFluideqns}. In particular, the terms containing the $b$-field are unambiguously associated to the spatial inhomogeneities in the quantum state variable $\hat\varrho$ and these inhomogeneities are responsible for the quantum backreaction on the classical fluid flow in the current  model.
Furthermore, we emphasize again that the internal energy has been assumed here to depend only on the fluid density. While this is suitable for barotropic fluids, adiabatic flows may also be realized by letting ${\cal E}={\cal E}(D,S)$ depend also on the specific entropy $S$, which is then Lie transported as an additional scalar function. On this occasion, however, we will restrict to the barotropic case for the sake of simplicity.

\section{Main features and discussion\label{sec:MainFeat}}

\subsection{Mead connection and  the  stress tensor\label{sec:MeadConn}}
As we observed in \S\ref{sec:backr}, the quantum-classical von Neumann operator \eqref{vnop} plays the same role as its phase-space variant from \S\ref{sec:hVNop}. Remarkably, this operator is expressed in terms of a non-Abelian connection form. Defined on a $\mathcal{U}(\mathsf{H})$-bundle over the configuration manifold $Q=\Bbb{R}^3$, this quantity is usually given as $i[\hat\varrho,\nabla\hat\varrho]$, where the imaginary unit is inserted so that this differential one-form takes values in the space $\operatorname{Her}(\mathsf{H})$ of Hermitian operators on $\mathsf{H}$. For later convenience, here we will refer to the connection form
\beq\label{MeadConn}
\widehat{\bGamma}=\frac{i\hbar}2[\hat\varrho,\nabla\hat\varrho]
\eeq
as the \emph{Mead connection}. Indeed, up to a numerical factor, the same quantity first appeared in Mead's work on geometric phases in molecular systems \cite{Me92}, although its role has  not been much investigated so far. Within the context of mixed quantum-classical dynamics, a phase-space extension of the Mead connection made its first appearance in the original model \eqref{HybEq1}-\eqref{HybEq3}. Indeed, as shown in \cite{GBTr22,TrGB23LNCS}, equation \eqref{HybEq2} may be rewritten as $\boldsymbol{\cal X}=
\langle \widehat{P}|\bX_{\widehat{\cal H}}\rangle+\big( \langle\bX_{{\widehat{\cal H}}}\cdot\nabla| (fJ\widehat{\boldsymbol\Upgamma})\rangle-\langle f(J\widehat{\boldsymbol\Upgamma})\cdot\nabla|\bX_{{\widehat{\cal H}}}\rangle
\big)/f$, where $J^{jk}=\{z^j,z^k\}$, $\widehat{\boldsymbol\Upgamma}={i\hbar}[\widehat{P},\nabla_{\!\bz}\widehat{P}]/2$, and we recall $\bz=(\bq,\bp)$. As we will see in this section, the Mead connection appears again in the present fluid closure, where it is defined on the configuration space.

\begin{remark}[Mead connection in spin-orbit coupling]\label{soc} The connection \eqref{MeadConn} plays a pivotal role in spin-orbit coupling (SOC),  in semirelativistic quantum mechanics.  In  this case, the  Hamiltonian operator is $\widehat{\cal H}=m^{-1}|\hat\bp|^2/2+{V(\hat{\bx})}+\widehat{\cal H}_{SOC}$, where $ \hat{\bp}=-i\hbar\nabla$  and 
\[
\widehat{\cal H}_{SOC}=-\frac\hbar{4 m^2 \mathsf{c}^2}\,\widehat{\bsigma}\times\nabla V\cdot\hat{\boldsymbol{p}}.
\]
Following from the Dirac equation, the SOC operator $\widehat{ \mathcal{H} }_{SOC}$ represents the quantum counterpart of the  Thomas precession  \cite{Baym,Thomas}.
As usual, the  Pauli equation $i\hbar\partial_t\Psi=\widehat{\cal H}\Psi$ is obtained from the Dirac-Frenkel action principle $\delta\int_{t_1}^{t_2}\!\int\operatorname{Re}(i\hbar\Psi^\dagger\partial_t\Psi-\Psi^\dagger\widehat{\cal H}\Psi)\,\de t=0$, where $\Psi(\bx)$ is  $\Bbb{C}^2$-valued and satisfies $\int\Psi^\dagger\Psi\,\de^3x=1$. Following \cite{BBirula}, we exploit  the exact factorization ansatz \cite{abedi2012correlated}  $\Psi(\bx)=\chi(\bx)\phi(\bx)$. Here,  $\chi\in L^2(\Bbb{R}^3)$ and $\phi(\bx)\in\Bbb{C}^2$ satisfies $\langle\phi(\bx),\phi(\bx)\rangle=1$ for all $\bx\in\Bbb{R}^3$. Then, the Dirac-Frenkel  action principle becomes $\delta\int_{t_1}^{t_2}\!\int\operatorname{Re}
\big(i\hbar\chi^*\partial_t\chi+|\chi|^2\langle\phi,i\hbar\partial_t\phi\rangle - \langle\chi\phi,\widehat{\cal H}(\chi\phi)\rangle\big)\de^3 x\de t=0$, and the SOC integrand  reads
\[
\langle\chi\phi,\widehat{\cal H}_{SOC}(\chi\phi)\rangle=\frac\hbar{4 m^2 \mathsf{c}^2}\nabla V\times\langle\widehat{\bsigma}\rangle\cdot\operatorname{Re}(\chi^*\hat{\boldsymbol{p}}\chi+|\chi|^2\mathbf{A})+\frac\hbar{4 m^2 \mathsf{c}^2}|\chi|^2\langle\nabla V\times\widehat{\bsigma},\widehat{\boldsymbol\Gamma}
\rangle
,
\]
with $\mathbf{A}:=\langle\phi,-i\hbar\nabla\phi\rangle$.
Here, we have denoted $\hat\varrho=\phi\phi^\dagger$ so that the Mead connection   \eqref{MeadConn} appears explicitly in the last term. 
The occurrence of $\widehat{\boldsymbol\Gamma}$ in SOC does not seem to have appeared  in the literature and a more thorough discussion is left for other venues.
\end{remark}

As anticipated, the  operator  \eqref{vnop} can be expressed in terms of the Mead connection as
$
\widehat{\mathscr{D}}
=D\hat\varrho+\operatorname{div}(c\widehat{\bGamma}\times\nabla b)
$,
and we notice that the classical density and the quantum density matrix are given by suitable projections of $\widehat{\mathscr{D}}$, that is $D=\operatorname{Tr}\widehat{\mathscr{D}}$ and $\hat{\uprho}=\int\widehat{\mathscr{D}}\,\de^3q$, respectively. Notice that, unlike other  quantum-classical models widely popular in chemical physics \cite{Bondarenko,Kapral}, here both the classical and the quantum density are positive. In more generality, the  model in \eqref{QCeqs1} satisfies the  consistency criteria outlined at the beginning of \S\ref{sec:PSmod}. In particular, the third property therein involves both quantum unitary transformations and classical  transformations on the configuration space (that is, cotangent lifts on phase space). 
The equivariance properties possessed by the equations \eqref{QCeqs1} follow from the equivariance of their underlying variational principle, which in turn hinges on the properties \eqref{equiv2} possessed by the von Neumann operator $\widehat{\mathscr{D}}$. In addition, following Proposition 5.11 in \cite{GBTr21}, the relations \eqref{equiv2} naturally lead to 
\begin{equation}\label{maria} 
\frac{\de \bm}{\de t}=-\int\langle\widehat{\mathscr{D}},\nabla\widehat{H}\rangle\,\de^3q
\,,\qquad\qquad
i\hbar\frac{\de \hat{\uprho}}{\de t}=\int[\widehat{H},\widehat{\mathscr{D}}]\,\de^3q.
\end{equation} 
In the case of   quantum-classical dynamics involving an infinite-dimensional quantum subsystem, we write $\widehat{H}(\bq)={H}(\bq,\hat{\bx},\hat{\bp})$, with $[\hat{\bx},\hat{\bp}]=i\hbar\boldsymbol{1}$. Then, the relations above ensure momentum conservation for translation-invariant Hamiltonians such as $\widehat{H}(\bq)=M^{-1}\hat{p}^2/2+{V(\bq-\hat{\bx})}$.

The  equations \eqref{maria}  are  verified explicitly by rewriting   \eqref{QCeqs1}, after several rearrangements, as
\begin{equation}\label{stress_form1}
\begin{aligned}
&MD(\partial_t+\bu\cdot\nabla)\bu=-
\langle \widehat{\mathscr{D}},\de \widehat{H}\rangle
-
\operatorname{div}{\sf T},\\
&i\hbar D(\partial_t+\bu\cdot\nabla)\hat\varrho=[\widehat{H},\widehat{\mathscr{D}}]+\operatorname{div}\!\Big(c\nabla b\times[ \widehat{H},\widehat{\bGamma}]+\frac{i\hbar}2c\nabla b\times\big[\hat{\varrho},[\hat{\varrho},\nabla \widehat{H}]\big]\!\Big),
\end{aligned}
\end{equation} 
where we have introduced the stress tensor
\begin{equation}\label{stress_T} 
{\sf T}_{jk}=
\big({\sf p} -c\nabla b \cdot \big\langle \widehat{\bGamma}, \times \nabla \widehat{H}\big\rangle\big)\delta_{jk}+c\langle\widehat{\bGamma},\times\nabla \widehat{H}\rangle_j\partial_k b+c\langle( \nabla b\times\widehat{\bGamma})_j, \partial_k\widehat{H}\rangle+ c\langle(\nabla \widehat{H}\times\nabla b)_j,\widehat{\Gamma}_k\rangle.
\end{equation} 
One verifies that ${\sf T}$ is symmetric by using the argument that any matrix of the form $(\mathbf{a}\times\mathbf{b})\mathbf{c}^T+(\mathbf{c}\times\mathbf{a})\mathbf{b}^T+(\mathbf{b}\times\mathbf{c})\mathbf{a}^T$ is symmetric by the  Jacobi identity. As shown in Appendix \ref{appendixB}, the symmetry of ${\sf T}$ follows from the fact that $\langle {\widehat{ \varrho }},i\hbar\{ \hat{ \varrho }, \widehat{H}\}_b\rangle$ is invariant under rotations of $\nabla \widehat{H}$, $ \nabla \hat{ \varrho }$, and $ \nabla b$.
Notice that this property would fail if the evolution equation \eqref{bevol} was dropped and the backreaction field  $b$ was treated as a constant: in that case, the last  relation in \eqref{vars3} would be replaced by $\delta b=0$, so that the resulting stress tensor would fail to be symmetric. In addition,
we observe that the Mead connection plays a predominant role not only in the construction of the hybrid von Neumann operator, but also in the characterization of the fluid stresses.  Before concluding this section, we observe that the momentum equation can be further rearranged as
\begin{multline*}
MD(\partial_t+\bu\cdot\nabla)\bu=-D
\langle {\hat\varrho},\de \widehat{H}\rangle
-
\langle (c\widehat{\bGamma}\times\nabla b)\cdot\nabla,\nabla\widehat{H}\rangle
+\langle(\nabla\widehat{H}\times\nabla b)\cdot\nabla,c\widehat{\bGamma}\rangle
\\
-
\nabla(\mathsf{p}-c\langle \hat\varrho,\{\hat\varrho,\widehat{H}\}_b\rangle)-\operatorname{div}(c\langle \widehat{\bGamma},\times\nabla \widehat{H}\rangle (\nabla b)^T),
\end{multline*}
and, using the definition \eqref{MeadConn}, we notice certain similarities between the first line above and the expression of the original phase-space vector field \eqref{HybEq2}.

\subsection{Hamiltonian structure and cross helicity}\label{sec:HamStrCrHel}

Having characterized the quantum-classical fluid system, its stress tensor, and the geometric quantities appearing therein, in this section we will show how more insight can be obtained by looking at the underlying Hamiltonian structure. As usual, the latter is given in terms of a Poisson structure and a Hamiltonian functional which identifies the energy of the system. 

The Hamiltonian structure of the fluid system \eqref{QCeqs1} may be easily found by considering the action principle $\delta\int_{t_1}^{t_2}l\,\de t=0$ associated to \eqref{Tiziana4} and introducing the convenient variable $\tilde\rho=D\hat\varrho$.
 Then, we have $\delta\int_{t_1}^{t_2}\!\big(\int(\bm\cdot\bu+\langle\tilde\rho,i\hbar\hat{\upxi}\rangle)\,\de^3 q-h\big)\de t=0$, where the Hamiltonian functional reads
\beq\label{Hamiltonian1}
h(\bm,\tilde\rho,D,b,c)=\int \!\bigg(\frac{|\bm|^2}{2MD}+D{\cal E}(D)+\bigg\langle \tilde\rho,{\widehat{H}}+\frac{i\hbar}{D^2} c\{{ \tilde\rho},\widehat{ H}\}_b\bigg\rangle\bigg)\,\de^3q.
\eeq
Notice that here we have restored the internal energy in \eqref{Tiziana4} in order to retain pressure effects. The candidate Poisson structure is found by applying the  relation $\dot{f}=\{f,h\}$  and using the equations of motion associated to an  arbitrary Hamiltonian. This step leads to the   bracket 
\begin{align}\nonumber
\{f,k\}(\bm,\tilde\rho,D,b,c)=&\int\!\bm\cdot\left(\frac{\delta k}{\delta \bm}\cdot\nabla\frac{\delta f}{\delta \bm}-\frac{\delta f}{\delta \bm}\cdot\nabla\frac{\delta k}{\delta \bm}\right)\de ^3 q
-
\int \! c\operatorname{div}\!\left(\frac{\delta f}{\delta \bm}\frac{\delta k}{\delta c}-\frac{\delta k}{\delta \bm}\frac{\delta f}{\delta c}\right)\de ^3 q
\\
&
-
\int \!D\left(\frac{\delta f}{\delta \bm}\cdot\nabla\frac{\delta k}{\delta D}-\frac{\delta k}{\delta \bm}\cdot\nabla\frac{\delta f}{\delta D}\right)\de ^3 q
-
\int \! b\operatorname{div}\!\left(\frac{\delta f}{\delta \bm}\frac{\delta k}{\delta b}-\frac{\delta k}{\delta \bm}\frac{\delta f}{\delta b}\right)\de ^3 q
\nonumber
\\
&
-\int\left\langle \tilde\rho,i\hbar^{-1}\left[\frac{\delta k}{\delta \tilde\rho},\frac{\delta h}{\delta \tilde\rho}\right]+\frac{\delta f}{\delta \bm}\cdot\nabla\frac{\delta k}{\delta \tilde\rho}-\frac{\delta k}{\delta \bm}\cdot\nabla\frac{\delta f}{\delta  \tilde\rho}\right\rangle\de^3  q,
\label{SDP-LPB-EF}
\end{align}
which   is Lie-Poisson on the dual of the semidirect-product Lie algebra
$
{\big(\mathfrak{X}(\Bbb{R}^3)\,\circledS\,{\cal F}(\Bbb{R}^3,\mathfrak{u}({\sf H}))\big)}\newline\,\circledS\,\big({\cal F}(\Bbb{R}^3) \times {\cal F}(\Bbb{R}^3)\times\operatorname{Den}(\Bbb{R}^3)\big)
$. Here, $\mathfrak{X}(\Bbb{R}^3)$ denotes the space of vector fields on $\Bbb{R}^3$, while $\mathfrak{u}({\sf H})$ identifies the Lie algebra of skew-Hermitian operators on the quantum Hilbert space ${\sf H}$. We observe that this bracket structure is similar to that underlying the fully quantum hydrodynamic equations \eqref{final-D-eqn}, although in that case the second terms in the first and second line are absent. Lie-Poisson structures of the type \eqref{SDP-LPB-EF} have appeared in several instances over the years,  in the context of complex fluid models \cite{GBRa09}.

The bracket \eqref{SDP-LPB-EF} possesses the Casimir invariants  
\beq
C_1=\operatorname{Tr}\int\! D F(D^{-1}\tilde\rho)\de^3q
\qquad\text{ and }\qquad
C_2=\int\!D\Phi(b,c,\{\Lambda_n\})\,\de^3q,
\label{Casimirs1}
\eeq
where  
\[
\Lambda_n=(D^{-1}\nabla c\times\nabla b\cdot\nabla)^n\|D^{-1}\tilde\rho\|
\,,\qquad\ 
n=1,2,\dots 
\]
satisfies the transport equation $\partial_t\Lambda_n+\bu\cdot\nabla\Lambda_n=0$  \cite{Henyey},
while $\Phi$ and $F$  are an arbitrary scalar function and  an arbitrary matrix analytic  function, respectively. We remark that the Casimirs $C_2$ extend a similar class of invariants appearing in the dynamics of ferromagnetic fluids \cite{HoKu88}.  While here we treat $C_1$ and $C_2$ separately for convenience, we notice that they may be combined into only one functional of the type $\operatorname{Tr}\int\! D {\sf F}(b,c,\Lambda_n,D^{-1}\tilde\rho)\de^3q$.

We will now show that more insight may be obtained upon writing $\tilde\rho=D\psi\psi^\dagger$, in which case   $C_1$ becomes a trivial constant and  $\Lambda_n$ vanishes, so that $C_2=\int\!D\Phi(b,c)\,\de^3q$.
From the first two in \eqref{LtoE2}, we find that the evolution of $\psi(t)$ is  $\psi=({\sf U}\psi_0) \circ  \boldsymbol\upeta^{-1} $,
 so that the time derivative gives $\partial_t\psi=\hat\upxi\psi-\bu\cdot\nabla\psi$,
and thus 
\[
\langle\tilde\rho,i\hbar\hat\upxi\rangle=D\langle\psi,i\hbar\partial_t\psi\rangle-D\bu\cdot\mathbf{A}
,\qquad\ \text{ with }\qquad\, 
\mathbf{A}=\langle\psi,-i\hbar\nabla\psi\rangle
\]
being the \emph{Berry connection} \cite{berry1984quantal}.
With this substitution and the introduction of the canonical fluid momentum $\bM=\bm-D\mathbf{A}$, the phase-space Lagrangian \eqref{Tiziana4} becomes  $\int(\bM\cdot\bu+D\langle\psi,i\hbar\partial_t\psi\rangle)\,\de^3 q-{h}$. Also, upon retaining the internal energy term,  the Hamiltonian functional reads
\beq
h(\bM,\psi,D,b,c)=\int \!\bigg(\frac{|\bM+D\mathbf{A}|^2}{2MD}+D{\cal E}(D)+\big\langle \psi,D{\widehat{H}}\psi
+c\nabla b\cdot\nabla\widehat{H}\times(i\hbar\nabla+\mathbf{A})\psi
 \big\rangle\bigg)\,\de^3q.
\label{ham2}
\eeq
In this case, since $\delta\psi$ and $\delta\bM$ are  arbitrary, and $\bM=D(M\bu-\mathbf{A})$, the equations \eqref{QCeqs1} become 
\begin{align}
MD(\partial_t+\bu\cdot\nabla)\bu=& -\bE-\bu\times\bB-\nabla{\sf p}-D\nabla\langle\widehat{H}\rangle+
\langle\psi,i\hbar \nabla b\cdot\widetilde{\mathbf{F}}\times\nabla\psi\rangle\nabla c
\\
& +\operatorname{div}\langle\psi,i\hbar c\widetilde{\mathbf{F}}\times\nabla\psi\rangle\,\nabla b
\nonumber
\\
i\hbar D(\partial_t+\bu\cdot\nabla)\psi
=&\,  D\widehat{H}\psi+{i\hbar}c\bigg(\{\psi\psi^\dagger,\widehat{H}\}_b+\{\widehat{H},\psi\psi^\dagger\}_b-\frac{1}{2c}\big[\{c, \widehat{H}\}_b,\psi\psi^\dagger\big]\bigg)\psi,
\label{untangledeqns}
\end{align}
where we have denoted
\[
\bE=-\partial_t\mathbf{A}-\nabla\langle\psi,i\hbar\partial_t\psi\rangle,\qquad\quad 
\bB=\nabla\times\mathbf{A},\qquad\quad 
\widetilde{\mathbf{F}}=-\big({\nabla \widehat{H}}-\langle\psi,{\nabla \widehat{H}}\psi\rangle\big).
\]
Notice that the equations \eqref{QCeqs2} remain unchanged.
Besides the usual Lorentz force, well known in molecular dynamics \cite{agostini2016quantum}, we observe the presence of an extra hydrodynamic force involving the \emph{fluctuation force operator} $\widetilde{\mathbf{F}}$ which identifies the quantum fluctuations around the Hellman-Feynman average $\langle\widehat{\mathbf{F}}\rangle=-\langle\psi,{\nabla \widehat{H}}\psi\rangle$ \cite{Feynman2}. We remark that the last term in the momentum equation above is exactly the same as the last term in the first equation of \eqref{QCeqs1}, although the use of the conditional state vector $\psi$ now unfolds the occurrence of the fluctuation force. This occurrence  in the present fluid model does not come as a surprise, since the same quantity already appears in the  phase-space model \eqref{HybEq1}-\eqref{HybEq3}, as shown in \cite{GBTr22}. 

\rem{ 
The Poisson bracket associated to the Hamiltonian \eqref{ham2} is 
\begin{align}\nonumber
\{f,k\}(\bm,\psi,D,b)=&\,\frac1{2\hbar}\operatorname{Im}\int\!\frac1D\left\langle \frac{\delta f}{\delta \psi}\bigg|\frac{\delta k}{\delta  \psi}\right\rangle\de ^3 q+\int\!\bm\cdot\left(\frac{\delta k}{\delta \bm}\cdot\nabla\frac{\delta f}{\delta \bm}-\frac{\delta f}{\delta \bm}\cdot\nabla\frac{\delta k}{\delta \bm}\right)\de ^3 q
\\
&
-
\int \!D\left(\frac{\delta f}{\delta \bm}\cdot\nabla\frac{\delta k}{\delta D}-\frac{\delta k}{\delta \bm}\cdot\nabla\frac{\delta f}{\delta D}\right)\de ^3 q
-
\int \! b\operatorname{div}\!\left(\frac{\delta f}{\delta \bm}\frac{\delta k}{\delta b}-\frac{\delta k}{\delta \bm}\frac{\delta f}{\delta b}\right)\de ^3 q
.
\label{SDP-LPB-EF2}
\end{align}
This is a Poisson structure on the direct sum $
{\cal F}({\Bbb{R}^3},{\sf H})\oplus\big(\mathfrak{X}(\Bbb{R}^3)\,\circledS\,\big({\cal F}(\Bbb{R}^3)\times\operatorname{Den}(\Bbb{R}^3)\big)$. 
\comment{CT: I see that this bracket is not optimal. Here, we should really define ${\Upsilon=\sqrt{D}\psi}$, but that would take too much work and would not give a lot of insight. Right now, I would refrain from going there.}
}  
Importantly, the momentum equation above can be written in terms of the Lie-derivative  as $(\partial_t+\pounds_\bu)(M\bu-\mathbf{A})=-\nabla(\delta h/\delta D)+ D^{-1}(\delta h/\delta c)\nabla c+D^{-1}(\delta h/\delta b)\nabla b$. On the one hand, upon computing $\delta h/\delta b= \operatorname{div}\langle\psi,i\hbar c\widetilde{\mathbf{F}}\times\nabla\psi\rangle$ and $ \delta h/\delta c=\nabla b\cdot\langle\psi,i\hbar \widetilde{\mathbf{F}}\times\nabla\psi\rangle$ this leads to the following  circulation dynamics:
\[
\frac{\de}{\de t}\oint_{\boldsymbol{\gamma}}(M\bu-\mathbf{A})\cdot\de\bq=\oint_{\boldsymbol{\gamma}}\frac1D\Big(\operatorname{div}\langle\psi,i\hbar c\widetilde{\mathbf{F}}\times\nabla\psi\rangle\,\nabla b+\langle\psi,i\hbar \nabla b\cdot\widetilde{\mathbf{F}}\times\nabla\psi\rangle\nabla c\Big)\cdot\de\bq,
\]
where   $\boldsymbol{\gamma}=\boldsymbol{\upeta} \circ \boldsymbol{\gamma}_0$ is any loop moving with the Lagrangian fluid path $\boldsymbol{\upeta}$. 
The Lie-derivative form of the momentum equation reads equivalently as $D(\partial_t+\pounds_\bu)(M\bu-\mathbf{A})=-D\nabla(\delta h/\delta D) 
+
(\delta h/\delta b)\nabla b+(\delta h/\delta c)\nabla c
$.
Consequently,  taking the dot product with $\nabla c\times\nabla b$,
one obtains the \emph{cross helicity} invariant
\beq
C_3=\int(M\bu-\mathbf{A})\cdot\nabla c\times\nabla b\,\de^3q.
\label{Casimir2}
\eeq
Here, the name  follows from similar invariants appearing in the hydrodynamics of magnetized plasmas \cite{Calkin}. An even more similar expression arises in hybrid kinetic-fluid  models \cite{MoTaTr15}.

We have seen how the use of the conditional state vector $\psi$, while unfolding the role of the fluctuation force $\widetilde{\bf F}$, also allows for a systematic characterization of circulation and cross helicity.
All these Casimirs may be used, for example, for a Lyapunov stability study via the energy-Casimir method \cite{HoMaRaWe85}.

\subsection{Pure-dephasing systems\label{sec:PureDef}}

As an informative specialization of our fluid model \eqref{QCeqs1}-\eqref{QCeqs2}, in this section we consider a particular type of quantum-classical systems, recently studied in \cite{MaRiTr23}. In the fully quantum formulation, a \emph{pure-dephasing}  system is given by a Hamiltonian operator of the type $\widehat{\cal H}={H}_0(\hat{\bx},\hat{\bp})+{H}_I(\hat{\bx},\hat{\bp},\widehat{A})$, where  $\widehat{A}$ is an operator commuting with the canonical observables $(\hat{\bx},\hat{\bp})$. 
For the present discussion, it is convenient to make extensive use of the notation $\langle\widehat{A}\rangle=\langle\hat\varrho,\widehat{A}\rangle$. Also, for the sake of simplicity, here we will consider the case when $\widehat{A}$ is a given Pauli matrix $\widehat{\sigma}_k$ and the fully quantum Hamiltonian is given as
\beq\label{PDHam}
\widehat{\cal H}=\frac1{2M}|\hat\bp|^2 +V_0(\hat\bx) +{V}_I(\hat{\bx})\widehat{\sigma}_k,
\eeq
We will now present a comparison of results obtained from the study of pure-dephasing dynamics in quantum hydrodynamics, the Ehrenfest fluid model, and quantum-classical hydrodynamics. As we will see, the latter succeeds in retaining the quantum backreaction on the classical flow, which instead is lost in Ehrenfest dynamics.

In first place, we specialize the quantum hydrodynamics equations  to the Hamiltonian \eqref{PDHam}, so that upon replacing  $\widehat{H}=V_0\boldsymbol{1} +{V}_I\widehat{\sigma}_k$ in \eqref{final-D-eqn}, we have
\begin{align}
\begin{split}
&  M(\partial_t + \bu\cdot\nabla)\bu   = 
  - \nabla (V_Q+V_0)-\langle\widehat{\sigma}_k\rangle\nabla V_I -\frac{\hbar^2}{2MD}\partial_j\langle D \nabla\hat{\varrho},  \partial_j\hat{\varrho}\rangle
  ,
  \\
  &
     i\hbar D(\partial_t +\bu\cdot\nabla)\hat{\varrho} = DV_I[\widehat{\sigma}_k,\hat{\varrho}]+\frac{\hbar^2}{2M}{\rm div}(D[\hat{\varrho},\nabla\hat{\varrho}])
,\qquad\quad
\partial_t D+{\rm div}(D\bu)=0\,,
\end{split}    \label{final-D-eqn2}
\end{align}
where  we have rearranged the quantum evolution equation. We observe that the overall expectation value $\langle\!\langle\widehat{\sigma}_k\rangle\!\rangle=\int \!D\langle\widehat{\sigma}_k\rangle\,\de^3 r$ remains constant in time, thereby ensuring physical consistency. In particular, the initial condition $\langle\!\langle\widehat{\sigma}_k\rangle\!\rangle=0$ is preserved by the dynamics. Nevertheless, the same does not hold for the local expectation $\langle\widehat{\sigma}_k\rangle$, which indeed has nontrivial dynamics. This means that the quantum degrees of freedom feed back in the hydrodynamic flow via both last two force terms in the momentum equation. Thus, despite the  simplicity of our pure-dephasing Hamiltonian \eqref{PDHam}, the full quantum hydrodynamics equations remain rather challenging.

The situation changes drastically in the case of the Ehrenfest fluid model. Indeed, if we replace   $\widehat{H}=V_0\boldsymbol{1} +{V}_I\widehat{\sigma}_k$ in \eqref{EhrenfestFluideqns}, we have 
\[
M(\partial_t + \bu\cdot\nabla)\bu   = - \nabla V_0
-\langle\widehat{\sigma}_k\rangle \nabla V_I
,\qquad\ 
     i\hbar (\partial_t +\bu\cdot\nabla)\hat{\varrho} = V_I[\widehat{\sigma}_k,\hat{\varrho}]
     ,\qquad\ 
     \partial_t D+{\rm div}(D\bu)=0.
\]
In this case, it is clear that the initial condition $\langle\widehat{\sigma}_k\rangle=0$  is preserved by the dynamics. This means that the hydrodynamic flow decouples entirely from the quantum motion, so that the Ehrenfest  model fails to capture any quantum backreaction in pure-dephasing systems \cite{MaRiTr23}.

Finally, let us now consider the quantum-classical fluid model in \eqref{QCeqs1}-\eqref{QCeqs2}. If we replace $\widehat{H}=V_0\boldsymbol{1} +{V}_I\widehat{\sigma}_k$ in \eqref{QCeqs2}, we obtain
\begin{align*}
MD(\partial_t+\bu\cdot\nabla)\bu=&- \nabla \mathsf{p}- D\nabla V_0
-D\langle\widehat{\sigma}_k\rangle \nabla V_I
+\frac\hbar2 \big\langle\nabla(c\hat\varrho),i\{[\hat\varrho,\widehat{\sigma}_k],V_I\}_b\big\rangle\\
&
+\frac\hbar2 \big\langle\nabla\hat\varrho,{i} \{c[\hat\varrho,\widehat{\sigma}_k],V_I\}_b\big\rangle-\frac\hbar2 \operatorname{div}\langle\hat\varrho,ic \nabla[\hat\varrho,\widehat{\sigma}_k]\times\nabla V_I\rangle\nabla b
\\
i\hbar D(\partial_t+\bu\cdot\nabla)\hat\varrho=&\, DV_I\big[\widehat{\sigma}_k,\hat\varrho\big]+{i\hbar}c\big[\{V_I,[\widehat{\sigma}_k,\hat\varrho]\}_b,\hat\varrho\big]-\frac{i\hbar}2\{c, V_I\}_b\big[\big[\widehat{\sigma}_k,\hat\varrho\big],\hat\varrho\big].
\end{align*}
Upon noticing that $\big\langle\widehat{\sigma}_k,\big[\{V_I,[\widehat{\sigma}_k,\hat\varrho]\}_b,\hat\varrho\big]\big\rangle
=\{V_I,\|[\widehat{\sigma}_k,\hat\varrho]\|^2\}_b/2$, we see that, in the case  $\hat\varrho=\psi\psi^\dagger$,   standard properties of the Pauli matrices lead to $\|[\widehat{\sigma}_k,\hat\varrho]\|^2=2(1-\langle\widehat{\sigma}_k\rangle^2)$ and thus
$
D(\partial_t+\bu\cdot\nabla)\langle\widehat{\sigma}_k\rangle=\{V_I, c(1-\langle\widehat{\sigma}_k\rangle^2)\}_b$.
This dynamics is generally nontrivial, so that, unlike the  the Ehrenfest model and similarly to the quantum case, the initial value $\langle\widehat{\sigma}_k\rangle$ is not generally preserved in time. This means that the quantum evolution keeps feeding back into the classical flow. However, unlike quantum hydrodynamics,  the fluid flow decouples entirely in the case when $V_I$ is spatially constant, in agreement with the item 4) of the consistency criteria discussed in the second paragraph of \S\ref{sec:PSmod}. 

The present study has shown that the quantum-classical hydrodynamic model  \eqref{QCeqs1}-\eqref{QCeqs2} overcomes the problematic force cancelations inherited from  Ehrenfest dynamics. Indeed, in the new model the quantum backreaction persists  through the  force terms containing the backreaction field $b$, which then acquires a fundamental role. While the persistence of backreaction  is shared with quantum hydrodynamics,   the absence of  second- and third-order gradients in the  model equations \eqref{QCeqs1}-\eqref{QCeqs2} represents a substantial difference, which may lead to  important simplifications from the viewpoint of both numerical and functional analysis. 

Before concluding this section, we emphasize that, while  emerging as drastic simplifications of more realistic problems, pure-dephasing systems have been considered in a variety of different fields, from optical physics \cite{BlHuWaGiSc04}, to quantum chemistry \cite{ReSiSu96}, and solvation dynamics \cite{Hughes}. In the context of  solvation hydrodynamics, quantum-classical pure-dephasing Hamiltonians were considered in \cite{Hughes}, although in that case the purely classical potential $V_0$ is replaced by a nonlocal convolution of the solvent density $D$. While this takes the problem to a higher level of difficulty, none of the previous arguments  would change in the presence of nonlocal potentials.

\subsection{Invariant planar subsystems\label{sec:planar}}

In the search for a simplified version of the quantum-classical fluid equations \eqref{QCeqs1}-\eqref{QCeqs2}, it is instructive to look for a lower-dimensional invariant subsystem. While we observe that the quantum backreaction is lost if we specialize our model to one spatial dimension, the same is not true for the two-dimensional case. Indeed, if we restrict to consider a planar flow with  $\bu=(\bv,0)$ and $\bv\in\mathfrak{X}(\Bbb{R}^2)$, then $b(x,y,z,t)=\beta z$ identifies an exact solution of the second equation in \eqref{QCeqs2}. 
If all the other variables are restricted to depend only on the planar coordinates, then the equations \eqref{QCeqs1}-\eqref{QCeqs2} become
\begin{align}\nonumber
MD(\partial_t+\bv\cdot\nabla)\bv=& -\nabla{\sf p}
-\langle\hat\varrho,D\nabla\widehat{H}\rangle
+\hbar \big\langle\nabla(\tilde{c}\hat\varrho),i\{\hat\varrho,\widehat{H}\}\big\rangle
+\hbar \big\langle\nabla\hat\varrho,{i} \{\tilde{c}\hat\varrho, \widehat{H}\}\big\rangle,\\
\nonumber
i\hbar D(\partial_t+\bv\cdot\nabla)\hat\varrho=&\, \Big[D\widehat{H}+{i\hbar}\Big(\tilde{c}\{\hat\varrho,\widehat{H}\}+\tilde{c}\{\widehat{H},\hat\varrho\}-\frac{1}{2}\big[\{\tilde{c}, \widehat{H}\},\hat\varrho\big]\Big),\hat\varrho\Big],\\
\partial_t D+{\rm div}(D\bv)=&\ 0
\,,\qquad\quad
\partial_t \tilde{c}+\bv\cdot\nabla \tilde{c}=0\,,
\label{QCeqs1bis-incomp}
\end{align}
where we have denoted $\tilde{c}=\beta c$ and $\{A,B\}=\mathbf{e}_3\cdot\nabla A\times\nabla B=\partial_xA\partial_y B-\partial_yA\partial_x B$. Here, no confusion should arise with the canonical phase-space bracket used in earlier sections. 
As it stands, the only substantial difference between the above planar subsystem and the full 3D model \eqref{QCeqs1}-\eqref{QCeqs2} is that the backreaction field has now become merely a numerical parameter $\beta$, now incorporated in the field $\tilde{c}=\beta c$. Nevertheless, the backreaction  terms  persist in the momentum equation, with the exception of the vertical forces.

We notice that these equations are again Hamiltonian with the Poisson bracket given by \eqref{SDP-LPB-EF} without the last terms in the second line, and the Hamiltonian \eqref{Hamiltonian1} with the replacement $c\{\,,\,\}_b\to\tilde{c}\{\,,\,\}$ in the last term. This means that the functional $C_1$ in \eqref{Casimirs1} is still a dynamical invariant, while the second Casimir drops to $C_2=\int D\Phi(\tilde{c},\{\Lambda_n\})\,\de^3q$. In turn, upon considering the case $\hat\varrho=\psi\psi^\dagger$,
the third invariant \eqref{Casimir2} becomes $C_3=\int\!\Omega\Theta(\tilde{c})\,\de^3q$, where   $\Omega=\mathbf{e}_3\cdot\nabla\times(M\bv-\mathbf{A})=\omega+\hbar\operatorname{Im}\{\psi^\dagger,\psi\}$ is the canonical vorticity and $\Theta$ is an arbitrary function.
In particular, we have $\nabla\times\mathbf{A}=B\mathbf{e}_3$ with $B=\hbar\operatorname{Im}\{\psi^\dagger,\psi\}$, and the equations of motion \eqref{untangledeqns} resulting in the case $\hat\varrho=\psi\psi^\dagger$ reduce to
\begin{align}\nonumber
&M(\partial_t+\bv\cdot\nabla)\bv= -\bE - B\bv\times\mathbf{e}_3 -D^{-1}\nabla{\sf p}-\nabla\langle\widehat{H}\rangle
+D^{-1} \langle\psi,i\hbar\mathbf{e
}_3\cdot\widetilde{\mathbf{F}}\times\nabla\psi\rangle\nabla\tilde{c},
\\
&\nonumber
i\hbar(\partial_t+\bv\cdot\nabla)\psi
= \widehat{H}\psi+{i\hbar}\frac{\tilde{c}}D\Big(\{\psi\psi^\dagger,\widehat{H}\}+\{\widehat{H},\psi\psi^\dagger\}-\frac{1}{2}\big[\{\ln \tilde{c}, \widehat{H}\},\psi\psi^\dagger\big]\Big)\psi,
\\
&\label{daniela}
\partial_tD+\operatorname{div}(D\bv)=0\,,\qquad\qquad\partial_t\tilde{c}+\bv\cdot\nabla\tilde{c}=0.
\end{align}

In this case, we have the following circulation law:
\[
\frac{\de}{\de t}\oint_{\boldsymbol{\gamma}}(M\bv-\mathbf{A})\cdot\de\bq=  \oint_{\boldsymbol{\gamma}}\frac1D\langle\psi,i\hbar\mathbf{e
}_3\cdot\widetilde{\mathbf{F}}\times\nabla\psi\rangle\nabla\tilde{c}\cdot\de\bq,
\]
where   $\boldsymbol{\gamma}(t)$ is again an arbitrary planar loop  moving with the Lagrangian fluid flow.

In the attempt to further simplify our quantum-classical fluid equations, we now consider the case of an incompressible fluid flow. In this case, we have $\operatorname{div}\bv=0$ and the volume form $\de^3q$ in physical space is preserved in time. For example, this situation could apply to an incompressible fluid solvent  interacting with a  quantum solute molecule. Then, since   $\tilde{c}=\beta D$ (recall the discussion in \S\ref{sec:backr}),  the equations \eqref{daniela} become
\begin{align}\nonumber
&M(\partial_t+\bv\cdot\nabla)\bv= -\bE - B\bv\times\mathbf{e}_3 -\nabla{\sf p}
+ \beta \langle\psi,i\hbar\mathbf{e
}_3\cdot\widetilde{\mathbf{F}}\times\nabla\psi\rangle\nabla\ln D,
\\
&\nonumber
i\hbar(\partial_t+\bv\cdot\nabla)\psi
=  \widehat{H}\psi+{i\hbar}\beta\Big(\{\psi\psi^\dagger,\widehat{H}\}+\{\widehat{H},\psi\psi^\dagger\}-\frac{1}{2}\big[\{\ln D, \widehat{H}\},\psi\psi^\dagger\big]\Big)\psi,
\\
&
\partial_tD+\bv\cdot\nabla D=0.
\label{QCeqs1incomp}
\end{align}
%
Here, the pressure $\sf p$ is now a Lagrange multiplier enforcing the condition $\operatorname{div}\bv=0$. We recognize that, even in the incompressible two-dimensional case, the backreaction forces persist  and this is true also for pure-dephasing systems. 
Similarly to the  case treated in the previous paragraph, the system
\eqref{QCeqs1incomp} possesses  the Casimir invariant  ${\cal C}=\int(\Omega\Theta(D)+\Phi(D))\,\de^3q$.

\section{Conclusions and perspectives}

The idea of restoring trajectories in quantum dynamics is extremely inviting, as it offers the possibility of new convenient computational schemes borrowing methods from classical simulation codes. However, despite the considerable work carried out over the last few decades on Madelung's hydrodynamics, the latter posses severe difficulties that continue to challenge the community. In this scenario, mixed quantum-classical models represent a promising perspective. While most of these models continue to be used in several test cases with a certain success, their underlying equations suffer from well-known consistency issues. The possible violation of Heisenberg's uncertainty principle in  the most popular method \cite{Bondarenko} is only one example \cite{AgCi07}.

Having  proposed a new Hamiltonian phase-space model overcoming these issues, here we have dealt with the problem of formulating a Hamiltonian fluid closure which goes beyond Ehrenfest dynamics and yet satisfies its consistency properties. As we showed, the proposed complex fluid model succeeds in capturing quantum-classical correlations in the case of pure-dephasing systems, which is precisely where the Ehrenfest model fails. Certain pure-dephasing systems were shown to be challenging also for other models alternative to  Ehrenfest \cite{BoGBTr19}.
Importantly, the proposed model has also the advantage of involving only first-order gradients. This  point represents a major simplification over Madelung hydrodynamics, whose main difficulties arise from the appearance of higher-order gradients in its equations of motion. This simplification happens at the expense of introducing two extra scalar fields that are transported by the flow. While the field $c$ is naturally linked to the fluid density, the backreaction field $b$ deserves further attention. Indeed,  it is not clear how $b$ should be initialized. One has a few possibilities in this regard.

A possible way to initialize $b$ consists in adopting a further closure, that is expressing it as a transported scalar that is constructed from the remaining variables. For example, the quantity $\operatorname{Tr} F(\hat\varrho)$ is transported for any matrix analytic function $F$ and one may set $b=\operatorname{Tr} F(\hat\varrho)$ for some $F$. For example, one may think of  setting $b=\alpha\operatorname{Tr} (\hat\varrho\ln\hat\varrho)$, for some parameter $\alpha$. The use of entropy functionals in hydrodynamic closures of mixed quantum-classical dynamics was first proposed in \cite{Bousquet}. More simply, one may also set $b=\|\hat\varrho\|^2$. However, if $\hat\varrho$ is initialized as a projection, these expressions reduce to a  constant thereby eliminating the backreaction. A possible alternative is to initialize $b$ by its equilibrium value $b_e$. Simple equilibria of the model equation \eqref{QCeqs1}-\eqref{QCeqs2} may be easily found by setting $\delta(h+C)=0$, where $C$ is one of the Casimirs treated in \S\ref{sec:HamStrCrHel}, or any combination thereof. For example, take the second in \eqref{Casimirs1} with the particular choice $C=\int D\Phi(b)\,\de^3 q$, where $\Phi$ is an arbitrary function. Since we are interested in the equilibria of $b$, we can simply set $\delta(h+C)/\delta b=0$. Absorbing the constants into $\Phi$, we obtain $\Phi'(b_e)=\operatorname{div}\langle c_e\hat\varrho_e,i \nabla\hat\varrho_e\times\nabla\widehat{H}\rangle/D_e$. Then, assuming that $\Phi'$ is invertible, we find that $b_e$ must be a function of the quantity $\operatorname{div}\langle c_e\hat\varrho_e,i \nabla\hat\varrho_e\times\nabla\widehat{H}\rangle/D_e$, thereby characterizing a possible initial profile of the backreaction field.

As we have seen, interesting connections to spin-orbit coupling  emerge and this is a completely unexplored direction suggesting that quantum-classical coupling may be modeled by the same algebraic structures appearing in the semirelativistic limit of the Dirac equation. Then, a natural open question concerns the comparison between the fully quantum  treatment of spin-orbit coupling in quantum hydrodynamics and the analogue treatment in mixed quantum-classical dynamics. Addressing this question requires extending the present model to the case in which the coupling depends on the momentum. This extension is currently under development.

Finally, one may wonder about possible numerical implementations of the proposed model. Given the simpler level of difficulty, we plan to start our computational efforts by focusing on the planar subystems. As the backreaction field is absorbed into a constant number,  not only does this case allow for less computational resources, but also  eliminates the necessity of dealing with the question about the initial profile of $b$. Given the presence of several advection equations and a variational formulation, this case would also offer a testing ground for recent structure-preserving finite-element schemes developed within our groups for the long-time simulation of fluid problems \cite{GaGa21}.

\paragraph{Acknowledgments.} We are grateful to Irene Burghardt, Francesco Di Maiolo, and Darryl Holm   for their keen remarks during the development of this work. We are also indebted with Paul Bergold for his careful reading of the manuscript.
This work was made possible through the support of Grant 62210 from the John Templeton Foundation. The opinions expressed in this publication are those of the authors and do not necessarily reflect the views of the John Templeton Foundation. Also, we acknowledge financial support by the Leverhulme Trust Research Project Grant RPG-2023-078 and the Royal Society Grant IES\textbackslash R3\textbackslash203005.

\addtocontents{toc}{\protect\setcounter{tocdepth}{0}}

\appendix

\section{Calculational details on the model equations}\label{appendixA}

We give here some details on the derivation of the quantum-classical fluid equations \eqref{QCeqs1} from the variational principle $ \delta \int_{t_1}^{t_2}\ell {\rm d} t=0$ with variations \eqref{vars2_5}--\eqref{vars3}.
First, we note that for an arbitrary Lagrangian $\ell(\boldsymbol{u},D,\hat{\upxi},\hat{\varrho},b)$, this action principle yields the general system
\begin{equation}\label{general_equations}
\begin{aligned} 
&\frac{d}{dt} \frac{\delta \ell}{\delta \bu}+ \pounds _\bu \frac{\delta \ell}{\delta \bu} = D \nabla  \frac{\delta \ell}{\delta D} -  \Big\langle \frac{\delta \ell}{\delta \hat{\upxi}}, \nabla \hat{\upxi} \Big\rangle - \frac{\delta \ell}{\delta b} \nabla b - \frac{\delta \ell}{\delta c} \nabla c - \Big\langle \frac{\delta \ell}{\delta \hat{\varrho}} , \nabla \hat{\varrho} \Big\rangle\\
&\frac{d}{dt}  \frac{\delta \ell}{\delta \hat{\upxi}} + \left[  \frac{\delta \ell}{\delta \hat{\upxi}} , \hat{\upxi}\right]  + \operatorname{div} \left(\frac{\delta \ell}{\delta \hat{\upxi}}\bu \right)  + \left[ \hat{\varrho}, \frac{\delta \ell}{\delta \hat{\varrho}} \right] =0.
\end{aligned}
\end{equation}  
This system is complemented by the advection equations for $D$, $b$, and $c$ given in \eqref{QCeqs2} as well as by the  equation for $\hat\varrho $  as in the second of \eqref{auxeqns}.
We now consider the Lagrangian \eqref{Tiziana5}.
In order to simplify the derivation, we note that the last integrand can be written as $\big\langle {\widehat{ \varrho }},i\hbar \boldsymbol{\beta}  \cdot \nabla  \hat{ \varrho } \times \nabla \widehat{H}\big\rangle$, for the one-form $ \boldsymbol{\beta}   = c \nabla b$. Hence the first equation can be equivalently written as
\begin{equation}\label{equation_beta_diamond} 
\frac{d}{dt} \frac{\delta \ell}{\delta \bu}+ \pounds _\bu \frac{\delta \ell}{\delta \bu} = D \nabla  \frac{\delta \ell}{\delta D} -  \left\langle \frac{\delta \ell}{\delta \hat{\upxi}}, \nabla \hat{\upxi} \right\rangle - \frac{\delta \ell}{\delta  \boldsymbol{\beta}  } \times \operatorname{curl} \boldsymbol{\beta}   + \boldsymbol{\beta}\operatorname{div} \frac{\delta \ell}{\delta  \boldsymbol{\beta}  }      - \left\langle \frac{\delta \ell}{\delta \hat{\varrho}} , \nabla \hat{\varrho} \right\rangle 
\end{equation}
which is obtained with the same variational principle as above, but written in terms of $ \boldsymbol{\beta}  $, with variations $ \delta \boldsymbol{\beta}  =      {\bw \times \nabla\times\boldsymbol{\beta }} - \nabla ( \boldsymbol{\beta} \cdot \bw)$.
The functional derivatives are ${\delta \ell}/{\delta \bu}=MD \bu$, ${\delta \ell}/{\delta \hat{\upxi}}=- i \hbar D \hat{\varrho}$, and ${\delta \ell}/{\delta \boldsymbol\beta }=-  \langle \hat \varrho , i \hbar \nabla  \hat{\varrho} \times \nabla \hat H  \rangle$, as well as
\begin{align*}
\frac{\delta \ell}{\delta D} &= \frac{M}{2} | \bu | ^2 - \mathcal{E} (D) - D \mathcal{E} '(D) + \langle \hat \varrho \mid i \hbar  \hat\upxi  \rangle -  \langle \hat \varrho \mid \hat H  \rangle \\
\frac{\delta \ell}{\delta \widehat{\varrho }}
&= i\hbar D\hat{\upxi}-D{\widehat{H}} - \frac{1}{2} ic\hbar\{\hat{ \varrho } , \hat{H} \}_b -  \frac{1}{2} ic\hbar \{\hat{H} ,\hat{ \varrho }\}_b + \partial _i\Big( \frac{i\hbar}{2} [ \nabla  \hat{H}, {\widehat{ \varrho }} ] \times c\nabla   b \Big) _i \,.
\end{align*} 
Noting the identities
\begin{align*} 
\partial _i\Big( {i\hbar} [ \nabla  \hat{H}, {\widehat{ \varrho }} ] \times c\nabla   b \Big) _i &=-{i\hbar}c \nabla b \cdot [ \nabla {\widehat{ \varrho }} , \times \nabla \hat H   ]  +  {i\hbar}  [ \{c, \hat H\}_b,  {\widehat{ \varrho }} ] \\
&= -{i\hbar} \{ \widehat{ \varrho } , \hat H\}_b -{i\hbar} \{ \hat H, \widehat{ \varrho }\}_b+  {i\hbar}  [ \{c, \hat H\}_b,  {\widehat{ \varrho }} ],
\end{align*} 
we can write
$
{\delta \ell}/{\delta \widehat{\varrho }}= i\hbar D\hat{\upxi}-D{\widehat{H}}- i\hbar c \{\hat{ \varrho } , \hat{H} \}_b -  i\hbar c\{\hat{H} ,\hat{ \varrho }\}_b+  {i\hbar}  [ \{c, \hat H\}_b,  {\widehat{ \varrho }} ]/2
$. When inserted in \eqref{equation_beta_diamond} we get
\begin{align*} 
D ( \partial _t \bu + \bu \cdot  \nabla \bu ) = &D \nabla \big( - \mathcal{E} (D) - D \mathcal{E} '(D) +  \langle \hat \varrho \mid i \hbar  \hat\upxi  \rangle -  \langle \hat \varrho \mid \hat H  \rangle\big)\\
& +   \langle i \hbar D \hat{\varrho}, \nabla \hat\upxi  \rangle +   \langle \hat \varrho , i \hbar \nabla  \hat{\varrho} \times \nabla \hat H  \rangle\times \operatorname{curl}  \boldsymbol{\beta}  - \boldsymbol{\beta}    \operatorname{div}  \langle \hat \varrho , i \hbar \nabla  \hat{\varrho} \times \nabla \hat H  \rangle\\
& -  \langle  i\hbar D\hat{\upxi}-D{\widehat{H}}- i\hbar c\{\hat{ \varrho } , \hat{H} \}_b -  i\hbar c\{\hat{H} ,\hat{ \varrho }\}_b+  \frac{i\hbar}{2}  [ \{c, \hat H\}_b,  {\widehat{ \varrho }} ] , \nabla \hat \varrho  \rangle \\
= &- \nabla \mathsf{p} -   \langle \hat \varrho \mid D\nabla \hat H  \rangle +   \langle \hat \varrho , i \hbar \nabla  \hat{\varrho} \times \nabla \hat H  \rangle\times \operatorname{curl}  \boldsymbol{\beta}  - \boldsymbol{\beta}    \operatorname{div}  \langle \hat \varrho , i \hbar \nabla  \hat{\varrho} \times \nabla \hat H  \rangle\\
& + \langle 2 i\hbar c \{\hat{ \varrho } , \hat{H} \}_b  -  \frac{i\hbar}{2}  [ \{c, \hat H\}_b,  {\widehat{ \varrho }} ] , \nabla \hat \varrho  \rangle.
\end{align*} 
The first equation in \eqref{QCeqs1} is then obtained by noting the following two equalities
\[
\langle\hat\varrho,i \hbar \nabla\hat\varrho\times\nabla\widehat{H}\rangle\times(\nabla c\times\nabla b)- c\nabla b\operatorname{div}\langle\hat\varrho,i \hbar \nabla\hat\varrho\times\nabla\widehat{H}\rangle = \hbar\langle\hat\varrho,i  \{\hat\varrho,\widehat{H}\}_b\rangle \nabla c- \hbar \operatorname{div}\langle c\hat\varrho,i  \nabla\hat\varrho\times\nabla\widehat{H}\rangle\nabla b
\]
\[
\langle 2 i\hbar c \{\hat{ \varrho } , \hat{H} \}_b  -  \frac{i\hbar}{2}  [ \{c, \hat H\}_b,  {\widehat{ \varrho }} ] , \nabla \hat \varrho  \rangle=   \hbar \langle\nabla\hat\varrho,ic\{\hat\varrho,\widehat{H}\}_b
+{i} \{c\hat\varrho, \widehat{H}\}_b \rangle.
\]
The second equation in \eqref{QCeqs1} follows directly from the second in \eqref{general_equations} by using the expression of the functional derivatives and the advection equation for $D$. Note that the variable $\hat\upxi$ has been eliminated, and that the resulting equation for $\hat \varrho $ is compatible with its advection given in the second of \eqref{auxeqns}.

\section{Stress tensor calculations}\label{appendixB}

We provide here some details concerning the form of the right hand side of the momentum equation in \eqref{stress_form1}, which involves the sum of two terms: the von Neumann operator term $\langle \widehat{\mathscr{D}}, {\rm d} \widehat{H} \rangle$ and the stress tensor term $ \operatorname{div}\mathsf{T}$. We also explain the symmetry of $\mathsf{T}$ as emerging from the rotational invariance of the Lagrangian in terms of $ \nabla \hat{ \varrho }$, $ \boldsymbol{\beta} = c \nabla b$, $ \nabla \widehat{H}$. 

\medskip\noindent\textbf{Structure of the momentum equation.} The situation is best explained by considering  the following general form of Lagrangian
\begin{equation}\label{gen_Lagrangian} 
\ell(\boldsymbol{u},D,\hat{\upxi},\hat{\varrho}, \boldsymbol{\beta} )= \int \left[  \frac{M}{2} D|\bu| ^2 - \epsilon \big(\hat{\upxi}, D, \hat{ \varrho }, \nabla  \hat{ \varrho }, \boldsymbol{\beta} ,\hat H, \nabla \widehat{H}\big) \right] {\rm d} ^3q,
\end{equation} 
which contains \eqref{Tiziana5} as a particular case, with $ \boldsymbol{\beta}  = c \nabla b$ as before. For this class of Lagrangians, \eqref{equation_beta_diamond} becomes
\begin{align} \nonumber
{MD} ( \partial _t \bu + \bu \cdot \nabla \bu)&= \operatorname{div} \Big( \underbrace{- (D \frac{\partial \epsilon }{\partial D} - \epsilon ) \delta - \frac{\partial \epsilon }{\partial  \boldsymbol{\beta} } \otimes  \boldsymbol{\beta} - \Big\langle  \frac{\partial \epsilon }{\partial \nabla  \hat{ \varrho }} ,\otimes  \nabla \hat{ \varrho } \Big\rangle -  \Big\langle \frac{\partial \epsilon }{\partial \nabla  \hat{ H }}, \otimes  \nabla \hat{ H }\Big\rangle }_{=-\mathsf{T} }\Big)\\
& \qquad  + \Big\langle  \underbrace{\partial _i \frac{\partial \epsilon }{\partial \partial _i \widehat{H}} -\frac{\partial \epsilon }{\partial \widehat{H}}}_{=-\widehat{\mathscr{D}}= - {\delta \ell}/{\delta \widehat{H}} },\nabla \widehat{H} \Big\rangle,
\label{gen_equations}
\end{align}
which shows the occurrence of a stress tensor $\mathsf{T}$ and a von Neumann operator, defined in general by $\widehat{\mathscr{D}}={\delta \ell}/{\delta \hat H}$. For our case
\[
\epsilon (\hat{\upxi}, D, \hat{ \varrho }, \nabla  \hat{ \varrho }, \boldsymbol{\beta} , \hat H, \nabla \widehat{H})= D
\big(  \mathcal{E} (D)- \big\langle \hat{\varrho}\big|i\hbar \hat{\upxi}\big\rangle  + \big\langle \hat{\varrho}\big|{\widehat{H}}\big\rangle \big) +\big\langle {\widehat{ \varrho }},i\hbar \boldsymbol{\beta}  \cdot \nabla  \hat{ \varrho } \times \nabla \widehat{H}\big\rangle,
\]
one gets
\begin{align*}
D\frac{\partial \epsilon }{\partial D} - \epsilon &=D ^2 \frac{\partial \mathcal{E} }{\partial D} - \boldsymbol{\beta}  \cdot \big\langle \widehat{\bGamma}, \times \nabla \widehat{H}\big\rangle &  \left\langle \frac{\partial \epsilon}{\partial \nabla \widehat{ \varrho }}, \otimes \nabla \widehat{ \varrho } \right\rangle& = \left\langle  \nabla   \widehat{H} \times \boldsymbol{\beta}  , \otimes   \widehat{ \boldsymbol{\Gamma}}  \right\rangle\\
\frac{\partial \epsilon}{\partial \boldsymbol{\beta} }  \otimes \boldsymbol{\beta}  &=  \big\langle \widehat{\bGamma}, \times \nabla \widehat{H}\big\rangle  \otimes \boldsymbol{\beta}  & \left\langle \frac{\partial \epsilon}{\partial \nabla \hat{ H}}, \otimes \nabla \widehat{H} \right\rangle & = \left\langle \boldsymbol{\beta}  \times \widehat{\bGamma}, \otimes \nabla \widehat{H} \right\rangle
\end{align*}
so that the stress tensor $\mathsf{T}$ as defined in \eqref{gen_equations} gives the expression \eqref{stress_T}.

\medskip\noindent\textbf{Symmetry of the stress tensor.} We here show how the symmetry of $\mathsf{T}$ is related to the rotational invariance of the term $\big\langle {\widehat{ \varrho }},i\hbar c\{ \hat{ \varrho }, \widehat{H}\}_b\big\rangle= \big\langle {\widehat{ \varrho }},i\hbar \boldsymbol{\beta}  \cdot \nabla  \hat{ \varrho } \times \nabla \widehat{H}\big\rangle$ under the transformation $ \boldsymbol{\beta} _i \rightarrow \boldsymbol{\beta} _jR^j_i$,  $ \partial _i \widehat{ \varrho } \rightarrow \partial _j\widehat{ \varrho } R^j_i$, $ \partial _i \widehat{ H } \rightarrow \partial _j\widehat{ H } R^j_i$, for all $R \in SO(3)$. This again is more easily seen by using the general Lagrangian \eqref{gen_Lagrangian}. We consider the following $SO(3)$ invariance of $ \epsilon $:
\begin{equation}\label{covariance} 
\epsilon (\hat{\upxi}, D, \hat{ \varrho }, \partial _j   \hat{ \varrho } R^j_i , \boldsymbol{\beta}  _j  R^j_i, \hat H, \partial _j  \widehat{H}R^j_i)= \epsilon (\hat{\upxi}, D, \hat{ \varrho }, \partial _i   \hat{ \varrho },\boldsymbol{\beta} ,\hat H, \partial _i  \widehat{H}),
\end{equation} 
for all $ R \in SO(3)$. In continuum mechanics, such types of invariance are related to objectivity and material frame indifference \cite{MaHu1983}.
Taking the derivative with respect to $R$ at the identity in the direction $ \xi \in \mathfrak{so}(3)$, we have
\[
\Big( \frac{\partial \epsilon }{\partial \boldsymbol{\beta} _i} \boldsymbol{\beta} _j + \Big\langle  \frac{\partial \epsilon }{\partial \partial _i \hat{ \varrho } } ,\partial _j \hat{ \varrho } \Big\rangle + \Big\langle \frac{\partial \epsilon }{\partial \partial _i \hat{ H } },  \partial _j \hat{ H } \Big\rangle \Big) \xi ^j_i=0,
\]
for all $ \xi \in \mathfrak{so}(3)$. This is equivalent to the symmetry of $ \frac{\partial \epsilon }{\partial \boldsymbol{\beta}  } \otimes \boldsymbol{\beta} - \left\langle  \frac{\partial \epsilon }{\partial \nabla  \hat{ \varrho }} ,\otimes  \nabla \hat{ \varrho } \right\rangle-  \left\langle \frac{\partial \epsilon }{\partial \nabla  \hat{ H }}, \otimes  \nabla \hat{ H }\right\rangle$. Recalling the general definition of $\mathsf{T}$ in \eqref{gen_equations}, this is equivalent to the \makebox{symmetry of the  tensor $\mathsf{T}$.}

\end{document}